\documentclass[aps,prb,twocolumn,letterpaper]{revtex4}

\usepackage{amsmath,amssymb,amsfonts,upgreek}
\usepackage{bm}
\usepackage{booktabs}
\usepackage{dcolumn}
\usepackage{color}
\usepackage{graphicx}
\usepackage{setspace}
\usepackage{multirow}
\graphicspath{{./}}

\begin{document}

\title{Calculation of model Hamiltonian parameters for LaMnO$_3$ using
  maximally localized Wannier functions}

\date{\today}

\author{Roman Kov\'a\v{c}ik}
\affiliation{School of Physics, Trinity College Dublin, Dublin 2, Ireland}
\email{kovacikr@tcd.ie}
\author{Claude Ederer}
\affiliation{School of Physics, Trinity College Dublin, Dublin 2, Ireland}

\begin{abstract}
Maximally localized Wannier functions (MLWFs) based on Kohn-Sham
band-structures provide a systematic way to construct realistic,
materials specific tight-binding models for further theoretical
analysis. Here, we construct MLWFs for the Mn $e_g$ bands in
LaMnO$_3$, and we monitor changes in the MLWF matrix elements induced
by different magnetic configurations and structural distortions. From
this we obtain values for the local Jahn-Teller and Hund's rule
coupling strength, the hopping amplitudes between all nearest and
further neighbors, and the corresponding reduction due to the
GdFeO$_3$-type distortion. By comparing our results with commonly used
model Hamiltonians for manganites, where electrons can hop between two
"$e_g$-like" orbitals located on each Mn site, we find that the most
crucial limitation of such models stems from neglecting changes in the
underlying Mn($d$)-O($p$) hybridization.
\end{abstract}


\maketitle

\section{Introduction}

The theoretical description of complex transition metal oxides and
similar materials is very often based on effective tight-binding (TB)
models, i.e. a representation of the electronic structure within a
certain energy region in terms of localized atomic-like orbitals.
Simple TB models with a small number of orbitals can be used to study
the essential mechanisms governing complex physical behavior, such as
for example that found in the colossal magneto-resistive
manganites.\cite{Dagotto/Hotta/Moreo:2001,Lin/Millis:2008}

The electronic properties of manganites $R_{1-x}A_x$MnO$_3$ ($R$:
trivalent rare earth cation, $A$: divalent alkaline earth cation) are
often described within an effective ``two-band'' TB model, where
electrons can hop between the two $e_g$ levels on each Mn site. The
corresponding Hamiltonian typically also contains several local terms
describing the coupling of the $e_{g}$ states to the $t_{2g}$ core
spin, to the Jahn-Teller (JT) distortion of the oxygen octahedra,
and/or the electron-electron Coulomb repulsion. It has recently been
shown, that such a model (with parameters obtained partly from first
principles calculations and partly by fitting to experimental data) is
able to reproduce the basic structure of the phase diagram as a
function of doping and temperature found in manganite systems such as
La$_{1-x}$(Ca,Sr)$_x$MnO$_3$.\cite{Lin/Millis:2008}

An elegant and systematic way to obtain realistic (materials-specific)
TB models is the construction of maximally localized Wannier functions
(MLFWs) from the Kohn-Sham states calculated using density functional
theory (DFT).\cite{1997_marzari} DFT calculations are known to give a
realistic description of electronic structure for systems where
electronic correlation effects are not too
strong.\cite{Jones/Gunnarsson:1989,Martin:Book} Furthermore, for
materials where correlation effects are important, a Wannier
representation of the Kohn-Sham band structure can be used to define a
subset of orbitals (the ``correlated subspace''), which can then be
used as basis for a more elaborate treatment of correlation effects
beyond standard DFT. This is done for example in DFT+DMFT (DMFT =
dynamical mean-field theory)
calculations,\cite{Georges_et_al:1996,Anisimov_et_al:1997,Kotliar/Vollhardt:2004,Lechermann_et_al:2006}
which aim at an accurate quantitative description of materials where
electronic correlation cannot be ignored.

In this work we construct MLWFs corresponding to the Mn $e_g$ states
for LaMnO$_3$, the parent compound for many manganite systems, based
on DFT calculations within the generalized gradient approximation
(GGA). We calculate the real space Hamiltonian matrix elements in the
MLWF basis for different structural modifications and for different
magnetic configurations, and we compare the obtained results with
assumptions made in commonly used two band TB models.

Our analysis is closely related to earlier work presented in
Ref.~\onlinecite{2007_ederer}, which examined the validity of the two
band picture by fitting TB model parameters (including the hopping
between nearest and next-nearest neighbors) to the DFT band structure
obtained within the local density approximation (LDA). The approach
based on MLWFs used in the present work is less biased and more
generally applicable, and thus allows for a more systematic analysis
than the manual fitting of TB parameters discussed in
Ref.~\onlinecite{2007_ederer}. It is also well suited for the
construction of the correlated orbital subspace used for DFT+DMFT
calculations.\cite{Lechermann_et_al:2006}

This paper is organized as follows. In the following section we
describe the theoretical background of our work. Thereby,
Sec.~\ref{sec:lmo-model} summarizes the effective two band model that
is often used for a theoretical treatment of manganites,
Sec.~\ref{sec:mlwf} presents the definition of the MLWFs,
Sec.~\ref{sec:struc} describes the various structural modifications of
LaMnO$_3$ investigated throughout this work, and
Sec.~\ref{sec:comp-details} lists some of the calculational
details. The presentation of results starts with the case of the ideal
cubic perovskite structure in Sec.~\ref{ss:i}. The individual effects
of the staggered JT and the GdFeO$_3$-type distortions are then
presented in Secs.~\ref{ss:ii} and \ref{ss:iii}, respectively. This is
followed by the results for the combined distortion in
Sec.~\ref{ss:iv-v}, and the construction of a refined TB model and its
application to the full experimental structure of LaMnO$_3$ in
Sec.~\ref{ss:tb}. Finally, the most important results and conclusions
are summarized in Sec.~\ref{sec:summary}.

\section{Method and theoretical background}

\subsection{Effective two-band models for LaMnO$_3$}
\label{sec:lmo-model}

\begin{figure}
  \includegraphics[width=0.78\columnwidth]{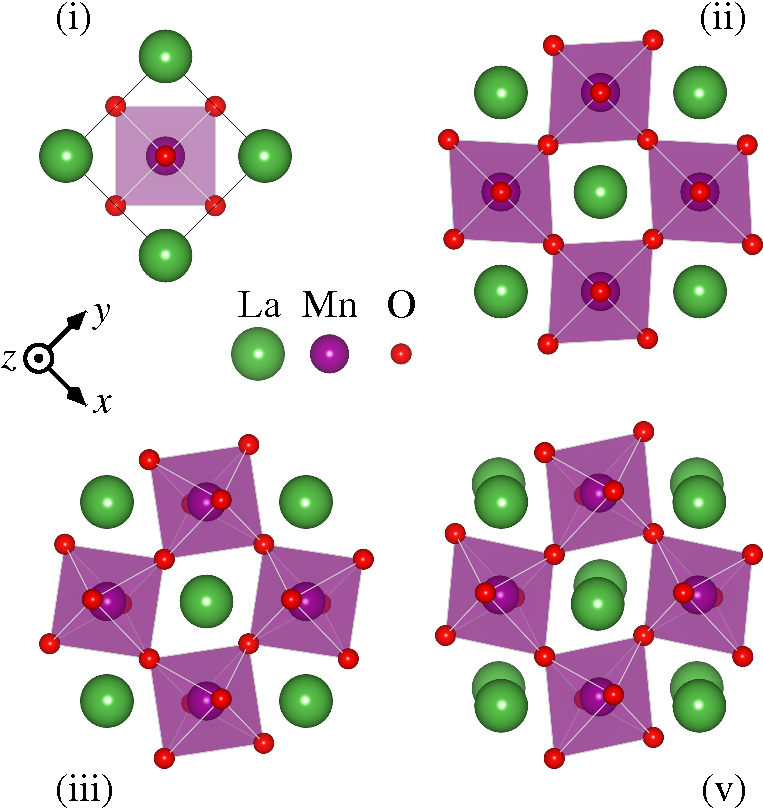}
  \caption{(Color online) Different structural modifications of
    LaMnO$_{3}$ investigated in this work, viewed along the [001]
    direction: (i)~ideal cubic perovskite, (ii)~purely Jahn-Teller
    distorted, (iii)~purely GdFeO$_{3}$-type distorted, and
    (v)~experimental $Pbnm$ structure. Pictures have been generated
    using VESTA.\cite{vesta}}
  \label{fig:lmo-cs}
\end{figure}

LaMnO$_{3}$ crystallizes in an orthorhombically distorted perovskite
structure with $Pbnm$ space group (see Fig.~\ref{fig:lmo-cs}v), and
A-type antiferromagnetic (A-AFM) order of the magnetic moments of the
Mn cations.\cite{Wollan/Koehler:1955,Elemans_et_al:1971} The deviation
from the simple cubic perovskite structure (Fig.~\ref{fig:lmo-cs}i)
can be decomposed into a staggered JT distortion of the MnO$_{6}$
octahedra within the $x$-$y$ plane (Fig.~\ref{fig:lmo-cs}ii), the
so-called GdFeO$_3$-type~(GFO) distortion, consisting of collective
tiltings and rotations of the oxygen octahedra
(Fig.~\ref{fig:lmo-cs}iii), and ``the rest'', i.e. displacements of
the La cations from their ideal positions plus a homogeneous
orthorhombic strain (Fig.~\ref{fig:lmo-cs}v).\cite{2007_ederer}

The electronic structure of LaMnO$_{3}$ close to the Fermi energy is
dominated by Mn 3$d$ states, which are split by the cubic component of
the crystal field into the lower-lying three-fold degenerate $t_{2g}$
and the higher-lying two-fold degenerate $e_{g}$
states.\cite{Pickett/Singh:1996,Satpathy/Popovic/Vukajlovic:1996,2007_ederer}
The formal electronic configuration \mbox{Mn$^{3+}$:~[Ar]~3$d^{4}$}
leads to a high spin state of the Mn cation with fully occupied local
majority spin $t_{2g}$ states and one electron per local majority spin
$e_{g}$ state, while both $t_{2g}$ and $e_{g}$ minority spin states
are empty.

Based on this electronic structure, the theoretical description of
manganites often employs an effective two-band TB picture, where
electrons can hop between the two $e_g$ levels on each Mn site. This
hopping is facilitated by hybridization with the oxygen 2$p$ states,
which, however, are not explicitly included in the TB model. It is
therefore understood, that the ``atomic'' $e_g$ states used in the TB
model are indeed somewhat extended Wannier orbitals that also include
the hybridization with the O 2$p$ states. In contrast, the three
$t_{2g}$ electrons are assumed to be tightly bound to a specific Mn
site where they give rise to a local ``core spin'' $S=3/2$. This core
spin then interacts with the valence $e_g$ electron spin via Hund's
rule coupling. In addition, a JT distortion of the surrounding oxygen
octahedron splits the two $e_g$ levels on the corresponding Mn site,
whereas elastic coupling between neighboring oxygen octahedra gives
rise to a cooperative effect. The GFO distortion in this picture is
usually assumed to simply reduce the effective hopping amplitudes
between neighboring Mn sites due to the resulting non-ideal Mn-O-Mn
bond angle. In addition, a local electron-electron interaction between
electrons occupying the same Mn site can be included in the
model.\cite{Dagotto/Hotta/Moreo:2001,Lin/Millis:2008}

The electronic Hamiltonian for such a model can be expressed as:
\begin{equation}
  \hat{H} = \hat{H}_\text{kin} + \hat{H}_\text{local} \ ,
\end{equation}
where
\begin{equation}
  \label{eq:TB-ham}
  \hat{H}_\text{kin} = \sum_{a,b,\mathbf{R},\Delta\mathbf{R},\sigma}
  t_{ab}(\Delta\mathbf{R}) \,
  \hat{c}^\dagger_{b(\mathbf{R}+\Delta\mathbf{R})\sigma}
  \hat{c}_{a\mathbf{R}\sigma} + \text{h.c.}
\end{equation}
describes the electron hopping between orbital $|a\rangle$ (spin
$\sigma$) at site $\mathbf{R}$ and orbital $|b\rangle$ at site
$\mathbf{R}+\Delta\mathbf{R}$, and it is assumed that all sites are
translationally equivalent, so that the hopping amplitudes
$t_{ab}(\Delta\mathbf{R})$ depend only on the relative position
between the two sites.

Representing the $e_g$ orbital subspace within the usual basis
$|1\rangle=|3z^2-r^2\rangle$ and $|2\rangle=|x^2-y^2\rangle$, and
assuming cubic symmetry, the nearest neighbor hopping along the three
cartesian directions has the following form:
\begin{align}
  \label{eq:hz}
  \mathbf{t}(\pm a_\text{c} \hat{\mathbf{z}}) & = \left(
    \begin{matrix}
      t & 0 \\
      0 & t'
    \end{matrix} \right)
  \\
  \label{eq:hx}
  \mathbf{t}(\pm a_\text{c} \hat{\mathbf{x}}) & = t \left(
    \begin{matrix}
      \tfrac14             & -\tfrac{\sqrt{3}}{4}\\
      -\tfrac{\sqrt{3}}{4} & \tfrac34
    \end{matrix} \right) + t' \left( \begin{matrix}
      \tfrac34             &  \tfrac{\sqrt{3}}{4}\\
      \tfrac{\sqrt{3}}{4} & \tfrac14
    \end{matrix} \right)
  \\
  \label{eq:hy}
  \mathbf{t}(\pm a_\text{c} \hat{\mathbf{y}}) & = t \left(
    \begin{matrix}
      \tfrac14             &  \tfrac{\sqrt{3}}{4}\\
      \tfrac{\sqrt{3}}{4} & \tfrac34
    \end{matrix} \right) + t' \left( \begin{matrix}
      \tfrac34             &  -\tfrac{\sqrt{3}}{4}\\
      -\tfrac{\sqrt{3}}{4} & \tfrac14
    \end{matrix} \right) \ .
\end{align}
Here, $a_\text{c}$ is the lattice constant of the underlying cubic
perovskite structure. The hopping $t'$ between two neighboring
$|x^2-y^2\rangle$-type orbitals along $\hat{\mathbf{z}}$ is small due
to the planar shape of this orbital, and it is therefore often
neglected. In this case, the nearest neighbor hopping depends only on
a single parameter $t$, the hopping along $\hat{\mathbf{z}}$ between
$|3z^2-r^2\rangle$-type orbitals.

$\hat{H}_\text{local}$ contains all local interaction terms included
in the model, i.e. Hund's rule coupling with the $t_{2g}$ core spin,
the JT coupling to the oxygen octahedra distortion, and eventually
also the electron-electron interaction. In this work we will discuss
only the Hund's rule and JT coupling, which are of the form:
\begin{equation}
\label{eq:hund}
  \hat{H}_\text{Hund} = -J \sum_{\mathbf{R}} \mathbf{S}_\mathbf{R} \cdot
  \mathbf{s}_\mathbf{R} \ , \ \text{and}
\end{equation}
\begin{equation}
\label{eq:jt}
  \hat{H}_\text{JT} = - \lambda \sum_{\mathbf{R},\sigma,a,b} \left(
    Q^x_\mathbf{R} \hat{c}^\dagger_{a\mathbf{R}\sigma} \tau^x_{ab}
    \hat{c}_{b\mathbf{R}\sigma} + Q^z_\mathbf{R} \hat{c}^\dagger_{a\mathbf{R}\sigma}
    \tau^z_{ab} \hat{c}_{b\mathbf{R}\sigma} \right) \ .
\end{equation}
Here, $J$ is the Hund's rule coupling strength and
$\mathbf{S}_{\mathbf{R}}$ is the $t_{2g}$ core spin at site
$\mathbf{R}$, which in the following we will consider as classical
vector normalized to
\mbox{$|\mathbf{S}_{\mathbf{R}}|=1$}. $\mathbf{s}_\mathbf{R}=\sum_{a,\sigma,\sigma'}
c^\dagger_{a\mathbf{R}\sigma} \bm{\tau}_{\sigma\sigma'}
c_{a\mathbf{R}\sigma'}$ is the corresponding $e_g$ valence spin,
$\lambda$ describes the strength of the JT coupling, and
$\bm{\tau}_{\sigma\sigma'}$ are the usual Pauli matrices. The
quantities $Q^x_\mathbf{R}$ and $Q^z_\mathbf{R}$ describe the JT
distortion of the oxygen octahedron surrounding site $\mathbf{R}$:
\begin{equation}
  Q^x_\mathbf{R} = \frac{1}{2\sqrt{2}} \left(d^x_{\mathbf{R}} -
    d^y_\mathbf{R} \right) \ , 
\end{equation}
\begin{equation}
  Q^z_\mathbf{R} = \frac{1}{2\sqrt{6}} \left( 2d^z_\mathbf{R} -
    d^x_\mathbf{R} - d^y_\mathbf{R} \right) \ ,
\end{equation}
where $d^x_\mathbf{R}$, $d^y_\mathbf{R}$, and $d^z_\mathbf{R}$
indicate the O-O distances along the $x$, $y$, and $z$ directions,
corresponding to the oxygen octahedron located at site $\mathbf{R}$.

\subsection{Maximally localized Wannier functions}
\label{sec:mlwf}

As is well known from basic solid state physics, the eigenfunctions
within a periodic crystal potential are extended Bloch waves,
classified by a wave-vector $\mathbf{k}$ and a band-index $m$. These
Bloch waves can in turn be expressed as a Bloch sum of ``atomic-like''
localized TB basis functions or \emph{Wannier functions}. For an
isolated group of $N$ Bloch states $\lvert \psi_{m\mathbf{k}}\rangle$,
i.e. a group of bands that are energetically separated from all lower-
or higher-lying bands throughout the entire Brillouin zone (BZ), a set
of $N$ localized Wannier functions $\lvert w_{n\mathbf{T}}\rangle$,
associated with lattice vector $\mathbf{T}$, is defined via the
following transformation:\cite{1997_marzari,2001_souza}
\begin{equation}\label{eq:wf}
  \lvert{w_{n\mathbf{T}}}\rangle= \frac{V}{\left({2\pi}\right)^{3}}
  \int_{\mathrm{BZ}} \left[{\sum_{m=1}^{N}
      U_{mn}^{\left(\mathbf{k}\right)}
      \lvert{\psi_{m\mathbf{k}}}\rangle} \right]
  \mathrm{e}^{-\mathrm{i}\mathbf{k}\cdot\mathbf{T}}\,\mathrm{d}\mathbf{k}
  \,.
\end{equation}
Here, $\mathbf{U}^{\left({\mathbf{k}}\right)}$ is a unitary matrix
mixing Bloch states at wave-vector $\mathbf{k}$. Different
$\mathbf{U}^{\left({\mathbf{k}}\right)}$ lead to different Wannier
orbitals, which are not uniquely determined by
Eq.~(\ref{eq:wf}). However, Marzari and Vanderbilt showed that a
unique set of \emph{maximally localized Wannier functions} (MLWFs) can
be obtained by minimizing the total quadratic spread of the Wannier
orbitals, defined as:\cite{1997_marzari}
\begin{equation}\label{eq:wfspread}
  \Omega = \sum_{n}^{N}\left[
    {\langle{r^2}\rangle_{n}-\langle\mathbf{r}\rangle_{n}^{2}} \right]
  \,,
\end{equation}
where $\langle \hat{O} \rangle_n = \langle w_{n\mathbf{0}} | \hat{O} |
w_{n\mathbf{0}} \rangle$.

For the case of entangled Bloch bands, i.e. bands that are not
energetically separated from other groups of higher- or lower-lying
states, an energy window $[E_\text{min},E_\text{max}]$ can be defined
such that there are
\mbox{$N_{\mathrm{win}}^{\left({\mathbf{k}}\right)}>N$} Bloch bands
within the energy window at each $\mathbf{k}$ vector, and then an
$N$-dimensional manifold of mixed Bloch states is obtained
as:\cite{2001_souza}
\begin{equation}\label{eq:wfdis}
  \lvert \psi_{m\mathbf{k}}^{\mathrm{dis}} \rangle = \sum_{l \in
    N_{\mathrm{win}}^{\left({\mathbf{k}}\right)}}
  U_{lm}^{\mathrm{dis}\left(\mathbf{k}\right)}
  \lvert{\psi_{l\mathbf{k}}}\rangle \,.
\end{equation}
The corresponding Wannier functions can then be obtained from the
mixed Bloch states by replacing $\lvert \psi_{m\mathbf{k}} \rangle$
with $\lvert \psi_{m\mathbf{k}}^{\mathrm{dis}} \rangle$ in
Eq.~(\ref{eq:wf}). The unitary rectangular
$N_{\mathrm{win}}^{\left({\mathbf{k}}\right)} \times N$ matrix
$\mathbf{U}^{\mathrm{dis}\left(\mathbf{k}\right)}$ is also uniquely
determined by the condition of maximal localization, i.e. it can be
obtained by minimizing
$\Omega\left(\mathbf{U}^{\left(\mathbf{k}\right)},
\mathbf{U}^{\mathrm{dis}\left(\mathbf{k}\right)}\right)$.\cite{2001_souza}

Once a set of MLWFs is obtained, the corresponding Hamilton matrix,
$\mathbf{H}^{\left(\mathrm{W}\right)}\left(\mathbf{k}\right)$, is
constructed by a unitary transformation:
\begin{equation}\label{hk}
  \mathbf{H}^{\left(\mathrm{W}\right)}\left(\mathbf{k}\right)=
  \bigl(\mathbf{U}^{(\mathbf{k})}\bigr)^{\dagger}
  \bigl(\mathbf{U}^{\mathrm{dis}(\mathbf{k})}\bigr)^{\dagger}
  \mathbf{H}^{(\text{B})}\left(\mathbf{k}\right)
  \mathbf{U}^{\mathrm{dis}(\mathbf{k})}
  \mathbf{U}^{(\mathbf{k})}
  \,,
\end{equation}
from the (diagonal) Hamilton matrix in the Bloch basis,
$H^{(\text{B})}_{nm}\left(\mathbf{k}\right)=\varepsilon_{n\mathbf{k}}\delta_{nm}$,
with eigenvalues $\varepsilon_{n\mathbf{k}}$. The MLWF Hamiltonian in
real space is then calculated as a Fourier transform of
$\mathbf{H}^{\left(\mathrm{W}\right)}\left(\mathbf{k}\right)$, which
in practice is replaced by a sum over $N_{k}$ points in
$\mathbf{k}$-space:
\begin{equation}\label{hr}
  h^\mathbf{T}_{nm} = \frac{1}{N_{k}}\sum_{\mathbf{k}}
  \mathrm{e}^{-\mathrm{i}\mathbf{k}\cdot\mathbf{T}}
  H_{nm}^{(\text{W})}\left(\mathbf{k}\right) \,.
\end{equation}
Thus, the real space representation of the Hamiltonian in the MLWF
basis is equivalent to a TB description of the full Hamiltonian within
the corresponding orbital subspace:
\begin{equation}
  \label{eq:tbh}
  \hat{H} = \sum_{\mathbf{T}, \Delta\mathbf{T}} h_{nm}^{\Delta\mathbf{T}}
  \, \hat{c}^\dagger_{n\mathbf{T}+\Delta\mathbf{T}} \hat{c}_{m\mathbf{T}}
  \ + \text{h.c.} \ ,
\end{equation}
where \mbox{$c_{m\mathbf{T}}$} is the annihilation operator for an
electron in orbital $\lvert w_{m\mathbf{T}} \rangle$. The real space
MLWF matrix elements $h_{nm}^{\mathbf{T}}$ can therefore be
interpreted as hopping amplitudes within a TB picture of MLWFs
[compare Eq.~(\ref{eq:tbh}) with Eq.~(\ref{eq:TB-ham})]. Note that
$\Delta\mathbf{T}$ in Eq.~(\ref{eq:tbh}) refers to lattice vectors,
whereas $\Delta\mathbf{R}$ in Eq.~(\ref{eq:TB-ham}) refers to Mn
sites. The subscripts $n$ and $m$ in Eq.~(\ref{eq:tbh}) can thus in
general indicate both site and orbital/spin character (for cases with
more than one site per unit cell).

For the case when MLWFs are constructed from an isolated set of bands,
the TB model, Eq.~(\ref{eq:tbh}), exactly reproduces the band
dispersion within the corresponding energy window. For the entangled
case, the energy bands calculated from Eq.~(\ref{eq:tbh}) do not
necessarily have to coincide with the underlying Bloch bands.

\subsection{Structural decomposition}
\label{sec:struc}

To analyze the effect of the various distinct structural distortions
within the experimental $Pbnm$ structure on the electronic properties
of LaMnO$_{3}$ we investigate several different atomic configurations
(similar to Ref.~\onlinecite{2007_ederer}):
\begin{enumerate}
\item[(i)] The ideal cubic perovskite
  structure~(Fig.~\ref{fig:lmo-cs}i).
\item[(ii)] A purely JT distorted structure (Fig.~\ref{fig:lmo-cs}ii),
  which results from alternating long and short O-O distances within
  the $x$-$y$ plane, i.e. a staggered JT distortion
  $Q^x_\mathbf{R}=\pm Q^x_\mathbf{0}$ and $Q^z_\mathbf{R}=0$. This
  distortion doubles the unit cell within the $x$-$y$ plane, leading
  to new in-plane lattice vectors
  \mbox{$\mathbf{a}_{\mathrm{ii}}=a_{\mathrm{c}}(\mathbf{\hat{y}}+\mathbf{\hat{x}})$}
  and
  \mbox{$\mathbf{b}_{\mathrm{ii}}=a_{\mathrm{c}}(\mathbf{\hat{y}}-\mathbf{\hat{x}})$}
  and tetragonal symmetry.
\item[(iii)] A purely GFO-distorted structure
  (Fig.~\ref{fig:lmo-cs}iii), resulting from rotations of the oxygen
  octahedra around the $z$ direction and octahedral tilts away from
  $z$, alternating along all three cartesian directions. This
  distortion quadruples the unit cell compared to the undistorted
  structure (i), yielding orthorhombic $Pbnm$ symmetry. The resulting
  in-plane lattice vectors are identical to those of structure (ii)
  and the new lattice vector along $z$ is
  \mbox{$\mathbf{c}_{\mathrm{iii}}=2a_{\mathrm{c}}\hat{\mathbf{z}}$}.
\item[(iv)] A superposition of JT and GFO distortion, which also leads
  to orthorhombic $Pbnm$ symmetry and unit cell vectors unchanged with
  respect to structure (iii).
\item[(v)] The full experimental structure (Fig.~\ref{fig:lmo-cs}v),
  with orthorhombically strained lattice vectors
  (\mbox{$|\mathbf{a}_{\mathrm{v}}| \neq |\mathbf{b}_\mathrm{v}| \neq
    |\mathbf{c}_\mathrm{v}|$}, resulting in \mbox{$Q^z_\mathbf{R} \neq
    0$}) and displaced La cations compared to structure (iv).
\end{enumerate}
For each of these structural modifications we use the same volume
\mbox{$V=60.91$~\AA$^3$} per formula unit as in the experimentally
observed $Pbnm$ structure.\cite{1995_norby} This leads to a cubic
lattice parameter $a_\text{c}=3.9345$~\AA, which deviates only by
\mbox{0.8$\,\%$} from the value we obtain by volume optimization for
the ideal perovskite structure within GGA. For the positions of the O
anions in structures (ii) and (iii) we use the same decomposition of
structure (iv) into pure JT and GFO components as described in
Ref.~\onlinecite{2007_ederer}~(see Table~\ref{tab:wyckpos}). For the
cases with A-AFM order, the unit cell is doubled in $z$ direction for
both (i) and (ii) structures in order to accommodate the magnetic
order, thus changing the symmetry to tetragonal in case (i).

\begin{table}
  \caption{Wyckoff parameters of the O(4c), ($x$, $y$, 0.25), O(8d),
    ($x$, $y$, $z$), and La(4c), ($x$, $y$, 0.25), sites for the
    various structural configurations used in this work (compare with
    Table I in Ref.~\onlinecite{2007_ederer}).}
  \label{tab:wyckpos}
  \begin{ruledtabular}
    \begin{tabular*}{\columnwidth}{@{\extracolsep{\fill}}lcccccc}
      &&Expt. (Ref.~\onlinecite{1995_norby})
      &(ii)&(iii)&(iv)&(v)\\
      \hline\\[-8pt]
      O(4c) &$x$&-0.0733& 0.0   &-0.0733&-0.0733&-0.0733\\
      &$y$&-0.0107& 0.0   &-0.0107&-0.0107&-0.0107\\
      \hline\\[-8pt]
      O(8d) &$x$& 0.2257& 0.2636& 0.2121& 0.2257& 0.2257\\
      &$y$& 0.3014& 0.2636& 0.2879& 0.3014& 0.3014\\
      &$z$& 0.0385& 0.0   & 0.0385& 0.0385& 0.0385\\
      \hline\\[-8pt]
      La(4c) &$x$& 0.0063& 0.0   & 0.0   & 0.0   & 0.0063\\
      &$y$& 0.5436& 0.5   & 0.5   & 0.5   & 0.5435\\
    \end{tabular*}
  \end{ruledtabular}
\end{table}

Starting from the ideal cubic perovskite structure, we analyze the
effect of a specific distortion by gradually increasing the amount of
this distortion, i.e. we perform series of calculations using a linear
superposition of the Wyckoff positions in the cubic perovskite
structure and in structure $(x)$:
\begin{equation}\label{eq:vardist}
\mathbf{R}(\alpha_x) = (1-\alpha_x)\,\mathbf{R}^{(\mathrm{i})} +
\alpha_{x}\,\mathbf{R}^{(x)} \,,
\end{equation}
and vary $\alpha_x$ between 0 and 1. The following cases are
considered: \mbox{($x= $ii)} (pure JT distortion), \mbox{($x= $iii)}
(pure GFO distortion), \mbox{($x= $iv)} (combined JT and GFO
distortions).

\subsection{Computational details}
\label{sec:comp-details}
We perform spin-polarized first principles DFT calculations using the
Quantum-ESPRESSO program package,~\cite{quantum-espresso} the GGA
exchange-correlation functional of Perdew, Burke, and
Ernzerhof,\cite{1996_perdew} and Vanderbilt ultrasoft
pseudopotentials~\cite{1990_vanderbilt} in which the
\mbox{La~(5$s$,5$p$)} and \mbox{Mn~(3$s$,3$p$)} semicore states are
included in the valence.

Convergence has been tested for the total energy and total
magnetization using the ideal cubic perovskite structure and
ferromagnetic (FM) order. We find the total energy converged to an
accuracy better than \mbox{1~mRy} and the total magnetization
converged to an accuracy of \mbox{0.05~$\mu_{\mathrm{B}}$} for a a
plane-wave energy cut-off of \mbox{35~Ry} and a $\Gamma$-centered
\mbox{${10}\!\times\!{10}\!\times\!{10}$} k-point grid using a
Gaussian broadening of \mbox{0.01~Ry}. These values for plane-wave
cutoff and Gaussian broadening are used in all calculations presented
in this work. The \mbox{${10}\!\times\!{10}\!\times\!{10}$} k-point
grid is used in all calculations for the cubic structure~(i), whereas
appropriately reduced k-point grids of
\mbox{${10}\!\times\!{10}\!\times\!{5}$},
\mbox{${7}\!\times\!{7}\!\times\!{10}$}, and
\mbox{${7}\!\times\!{7}\!\times\!{5}$} are used for the structures
with unit cell doubled in the $z$ direction, doubled in the $x$-$y$
plane, and quadrupled, respectively.

After obtaining the DFT Bloch bands within GGA, we construct MLWFs
using the \texttt{wannier90} program integrated into the
Quantum-ESPRESSO package.~\cite{Mostofi_et_al:2008} Starting from an
initial projection of the Bloch bands onto atomic $d$ basis functions
\mbox{$\lvert{3z^2-r^2}\rangle$} and \mbox{$\lvert{x^2-y^2}\rangle$}
centered at the different Mn sites within the unit cell, we obtain a
set of two $e_{g}$-like MLWFs per spin channel for each site.  The
spread functional (both gauge-invariant and non-gauge-invariant parts)
is considered to be converged if the corresponding fractional change
between two successive iterations is smaller than $10^{-10}$.  For
cases with entangled bands a suitable energy window is chosen as
described in the corresponding ``Results'' section.

\section{Results and Discussion}

\subsection{Perfect cubic perovskite -- structure~(i)}\label{ss:i}

\begin{figure}
  \includegraphics[width=\columnwidth]{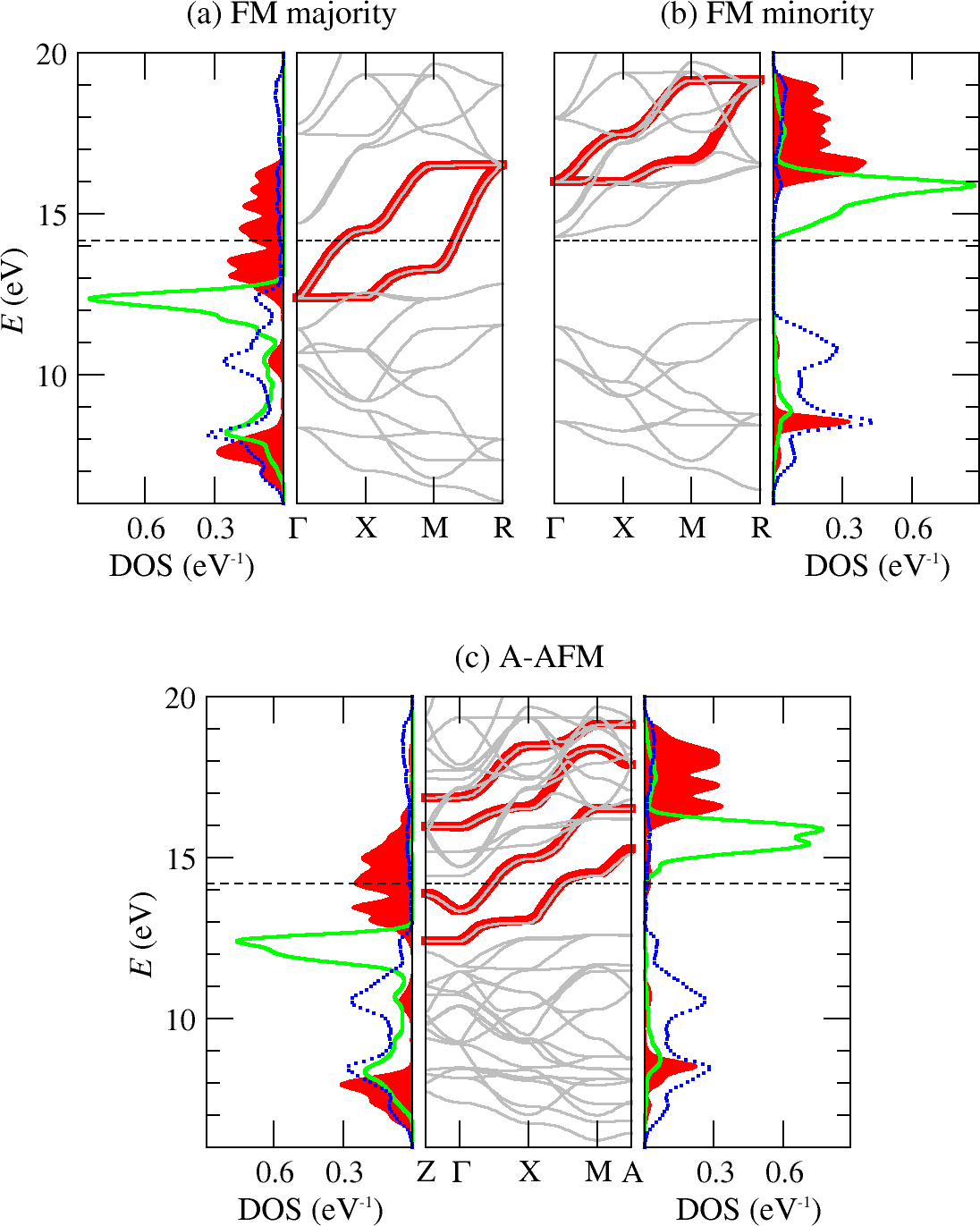}
  \caption{(Color online) Projected DOS and band structure along high
    symmetry lines within the BZ calculated for the cubic
    structure~(i) and both FM and A-AFM order.  Filled (red) areas and
    solid (green) lines represent the projected DOS corresponding to
    Mn($e_{g}$) and Mn($t_{2g}$) states, respectively, while dashed
    (blue) lines represent the site and orbitally averaged projected
    DOS corresponding to the O $p$ states. For the A-AFM case the left
    (right) panel corresponds to local majority (minority) spin
    projection. In the band structure plots, the dispersion calculated
    from the $e_{g}$-like MLWFs are represented by thick (red) lines.
    The Fermi level is indicated by the horizontal dashed lines.}
  \label{fig:c-pdos-bs}
\end{figure}

The projected densities of states~(DOS) and band structure calculated
for LaMnO$_3$ in the ideal cubic perovskite structure~(i) for both FM
and A-AFM order are shown in
Fig.~\ref{fig:c-pdos-bs}.\cite{footnote:k-points}

A metallic state is obtained for both FM and A-AFM order, in agreement
with previous band-structure
calculations.\cite{Pickett/Singh:1996,Satpathy/Popovic/Vukajlovic:1996,2007_ederer}
The projected DOS show that the (local) majority spin bands around the
Fermi energy have mainly Mn($e_{g}$) character and are half-filled
while the (local) minority spin bands with mainly Mn($e_{g}$)
character are unoccupied, as expected from the formal electron
configuration. Bands with Mn($t_{2g}$) character are lying just below
the Mn($e_{g}$) bands, and slightly overlap with the latter. O($p$)
bands are located below the Mn($t_{2g}$) bands (between
$\sim$\,6-12~eV) and are fully occupied.  The strong hybridization
between Mn($d$) and O($p$) electrons can be seen from the substantial
amount of Mn($d$) character in the energy range around \mbox{8~eV},
i.e. towards the bottom of the bands with predominant O($p$)
character. The states above the Mn($e_{g}$) bands have predominantly
La($d$) character.

One can see from the band structures depicted in
Fig.~\ref{fig:c-pdos-bs} that for the FM majority spin channel the
bands with predominant $e_g$ character are nearly completely isolated
from both higher and lower-lying bands, while for the FM minority spin
channel and in the A-AFM case, the ``$e_g$ bands'' overlap strongly
with other bands (with mostly Mn($t_{2g}$) minority and La($d$)
character). As described in section~\ref{sec:comp-details}, in order
to construct $e_g$-like MLWFs for the various cases, we define an
energy window for the disentanglement procedure [see
Eq.~(\ref{eq:wfdis})], and then initialize the Wannier functions from
a projection of the Kohn-Sham states within that energy window on
atomic $e_g$ wave-functions (see Ref.~\onlinecite{2001_souza}). A
suitable energy window is chosen based on the $e_g$ projected DOS and
calculated band structure (see discussion below for more details). Two
MLWFs per spin channel for the single Mn site within the cubic unit
cell are constructed for FM order, and two pairs of MLWFs, localized
at the two Mn sites within the magnetic unit cell, are constructed for
A-AFM order (for global spin up projection only).

\begin{figure}
  \includegraphics[width=0.9\columnwidth]{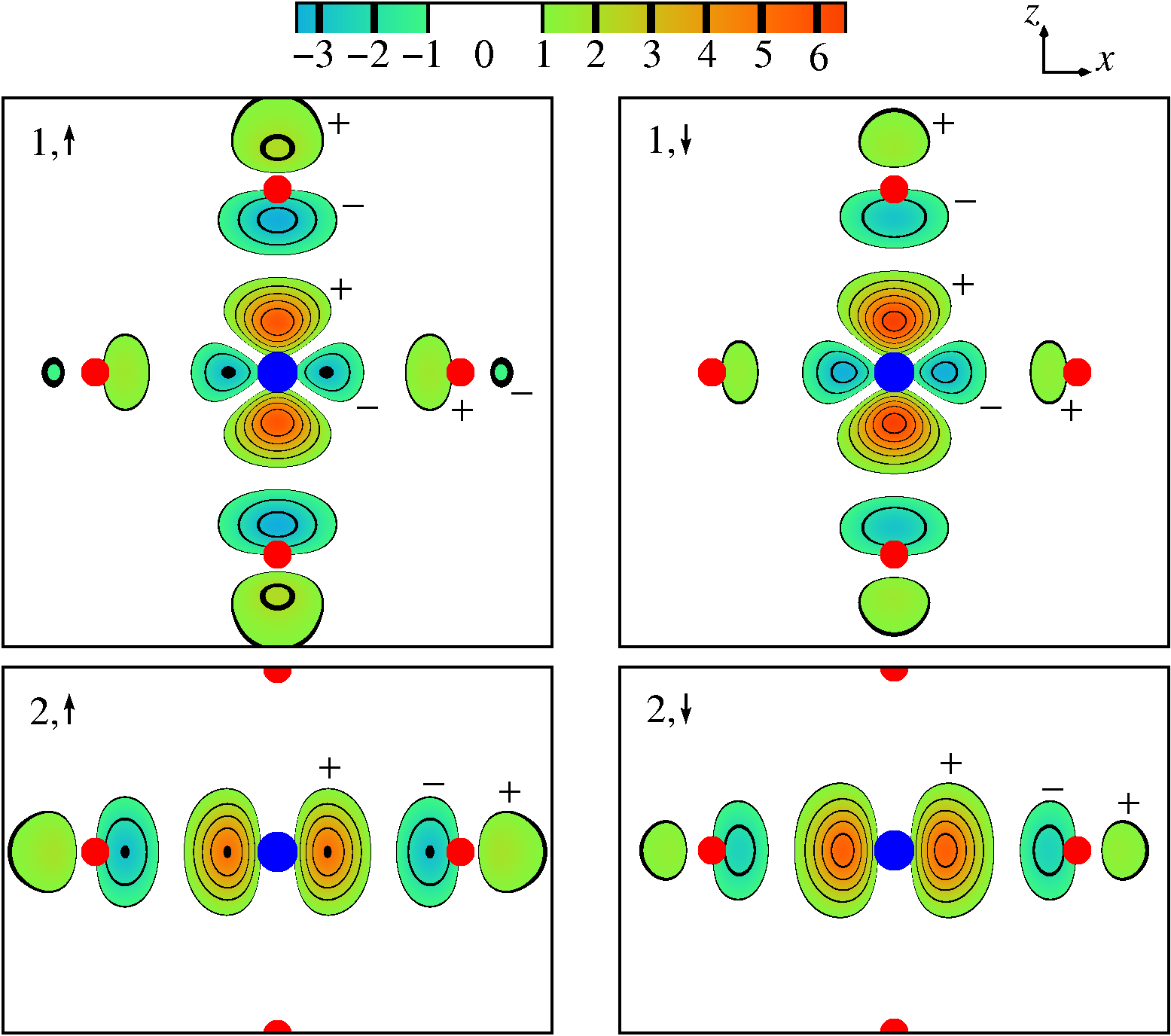}
  \caption{(Color online) Real space representation of the MLWFs for
    majority and minority spin projections in the cubic structure~(i)
    with FM order, projected on the $x$-$z$ plane passing through Mn
    (large/blue sphere) and O (small/red spheres) sites (in arbitrary
    units).}
  \label{fig:c-mlwf}
\end{figure}

Figure~\ref{fig:c-mlwf} shows the real space representation of the two
$e_{g}$-like MLWFs for both majority and minority spin and FM order,
calculated for an energy window of \mbox{[12.0, 17.0]~eV} and
\mbox{[15.9, 20.0]~eV}, respectively. The shape of the MLWFs resembles
the antibonding $\sigma^{*}$ character of hybridization between
Mn($e_{g}$) and O($p$) states in this energy range. The hybridization
is notably stronger for the majority spin MLWFs (individual spread per
WF \mbox{$2.90$~$\mathrm{\AA^{2}}$} compared to
\mbox{$1.65$~$\mathrm{\AA^{2}}$} for the minority spin MLWFs), which
is due to the smaller energy separation between the atomic Mn($e_{g}$)
and O($p$) levels for the majority spin channel. The difference
between the real space representation of the MLWFs for FM and A-AFM
order (not shown here) is more subtle. A quantitative comparison of
the corresponding differences in the real space Hamilton matrix
elements between MLWFs will be presented below.

The dispersion calculated from the obtained $e_g$ MLWFs is also shown
in Fig.~\ref{fig:c-pdos-bs}. It can be seen that even in the cases
with strongly entangled $e_g$ bands (FM minority spin and A-AFM) the
MLWF bands follow certain DFT bands almost exactly, except around some
band crossings with higher lying La $d$ bands. This represents the
fact that within cubic symmetry the $e_g$ states cannot hybridize with
the $t_{2g}$ bands, and hybridize only very weakly with the La $d$
states.

In order to reproduce the two majority spin bands around the Fermi
energy for the FM case, the lower bound of the energy window,
$E_\text{min}$, has to be above the lower peaks in the Mn($e_g$)
projected DOS at around \mbox{10.5~eV} and \mbox{8~eV}, which
correspond to the bonding combination of hybridized atomic O($p$) and
Mn($e_g$) states.  If these bands are included in the energy window,
the bonding and antibonding combinations of atomic orbitals become
disentangled and the $e_g$ Wannier functions become essentially
``atomic-like'' (compare also with the case of SrVO$_3$ described in
Ref.~\onlinecite{Lechermann_et_al:2006}). On the other hand, varying
$E_\text{min}$ within \mbox{0.4~eV} below the $\Gamma$ point energy of
the $e_g$-like bands changes the MLWF bands by less than \mbox{1~meV}
for any $\mathbf{k}$. Similarly, varying the upper bound of the energy
window has only minor influence on the resulting MLWF bands, due to
the negligible hybridization of the $e_g$ states with higher-lying
bands. Additional test calculations for different k-point grids showed
that the MLWF band structure is converged within \mbox{0.5~meV} at any
k-point for the \mbox{${10}\!\times\!{10}\!\times\!{10}$} grid which
was used for the energy window test calculations.

\begin{figure}
  \includegraphics[width=0.9\columnwidth]{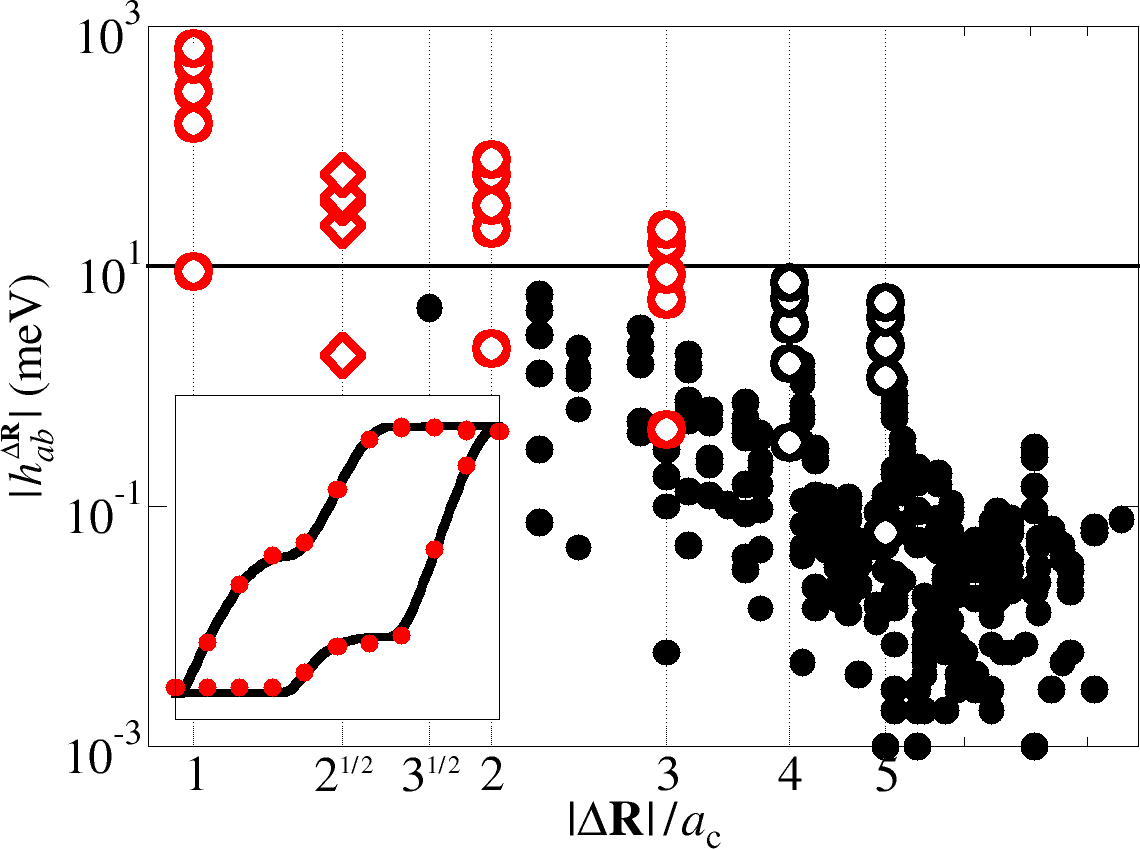}
  \caption{(Color online) Magnitude of all calculated non-zero hopping
    parameters for FM order in the ideal cubic structure as a function
    of the inter-site distance $|\Delta\mathbf{R}|$ (open circles:
    hopping along the unit cell directions; open diamonds: hopping
    between next-nearest neighbors; filled circles: all other
    hoppings). Inset: Comparison of the full MLWF band structure
    (solid lines) and the one calculated from a simplified TB model
    (filled circles) which includes only the inter-site hoppings for
    which the largest matrix element is larger than \mbox{10~meV} (see
    main text).}
  \label{fig:cfu-habsr}
\end{figure}

We now turn to the analysis of the hopping parameters, i.e. the real
space matrix elements $h_{ab}^{\Delta\mathbf{R}}$, Eq.~(\ref{hr}),
between MLWFs located at different Mn sites.  The magnitudes of all
calculated hopping parameters for the FM majority spin case are shown
in Fig.~\ref{fig:cfu-habsr}.  It is noticeable that the hopping
amplitudes along the three cartesian axes are most dominant and that
their decay with distance is rather slow, so that the terms
corresponding to inter-site distances of $2a_\mathrm{c}$ and
$3a_\mathrm{c}$ are of comparable magnitude as the hopping between
next-nearest neighbors for which
\mbox{$\lvert\Delta\mathbf{R}\rvert=\sqrt{2}a_{\mathrm{c}}$}.

The exact MLWF representation in terms of
$\mathbf{H}^{(\text{W})}(\mathbf{k})$ is well suited for further
numerical calculations, e.g. within a DFT+DMFT approach. On the other
hand, for the analysis of specific physical mechanisms within a
semi-analytical TB model, one generally wants to use only a very
limited number of hopping parameters $\mathbf{h}^{\Delta\mathbf{R}}$
between closest neighbors. We therefore identify a minimal subset of
hopping parameters, corresponding to intersite distances
\mbox{$\lvert\Delta\mathbf{R}\rvert/a_{\mathrm{c}}\in\{1,\sqrt{2},2,3\}$},
i.e. where only hopping between sites, for which the leading term
(i.e. the corresponding matrix element with largest magnitude) is
larger than \mbox{10~meV}, are considered, while the rest is set to
zero. This model yields an overall very good agreement with the full
MLWF band structure (see inset in Fig.~\ref{fig:cfu-habsr}), deviating
not more than \mbox{0.11~eV} for any k-point on the
\mbox{${10}\!\times\!{10}\!\times\!{10}$} k-point grid used. On the
other hand, a TB model where only the hopping amplitudes between
nearest and next-nearest neighbors are taken into account leads to
deviations of up to \mbox{0.29~eV} for some k-points, which might
still be acceptable for certain purposes. However, the overall
bandwidth for the latter model is reduced by about \mbox{0.2~eV}
compared to the full MLWF band structure.

\begin{table}
  \caption{Calculated values of the on-site, nearest, and next-nearest
    neighbor matrix elements $h_{ab}^{\Delta\mathbf{R}}$ (in meV) for
    FM and A-AFM order within structure~(i) for the two different spin
    projections. As described in the text, in the A-AFM case all
    matrix elements refer to the Mn site closest to the origin.}
  \label{tab:c-hp}
  \newcolumntype{.}{D{.}{.}{5.1}}
  \begin{ruledtabular}
    \begin{tabular*}{\columnwidth}{@{\hspace{2pt}}@{\extracolsep{\fill}}l....}
      (a)
      &\multicolumn{1}{c}{FM($\uparrow$)}
      &\multicolumn{1}{c}{FM($\downarrow$)}
      &\multicolumn{1}{c}{A-AFM($\uparrow$)}
      &\multicolumn{1}{c}{A-AFM($\downarrow$)}\\
      \hline\\[-8pt]
      $h_{11}^{0}$&14485.7&17483.9&14638.9&17379.3\\[2.5pt]
      $h_{22}^{0}$&14484.3&17483.7&14541.7&17443.3\\[1pt]
      \hline\\[-8pt]
      $h_{11}^{z}$&-648.2&-512.5&\multicolumn{2}{.}{-595.0}\\[2.5pt]
      $h_{22}^{z}$&9.1&-9.2&\multicolumn{2}{.}{-8.5}\\[2.5pt]
      $h_{11}^{x}$&-155.2&-135.0&-172.8&-130.7\\[2.5pt]
      $h_{12}^{x}$&284.5&217.9&281.4&214.5\\[2.5pt]
      $h_{22}^{x}$&-483.9&-386.7&-488.8&-389.8\\[1pt]
      \hline\\[-8pt]
      $h_{11}^{xz}$&37.8&32.0&\multicolumn{2}{.}{34.7}\\[2.5pt]
      $h_{12}^{xz}$&-34.1&-26.6&-29.3&-29.0\\[2.5pt]
      $h_{22}^{xz}$&-1.8&1.2&\multicolumn{2}{.}{0.5}\\[2.5pt]
      $h_{11}^{xy}$&-21.9&-14.3&-16.4&-14.2\\[2.5pt]
      $h_{22}^{xy}$&57.9&47.4&51.4&48.3\\
    \end{tabular*}
  \end{ruledtabular}
\end{table}

The calculated matrix elements of the real space matrix elements
$h_{ab}^{\Delta\mathbf{R}}$ for nearest and next nearest neighbor
hopping as well as the corresponding on-site terms
($\Delta\mathbf{R}=0$) are summarized in Table~\ref{tab:c-hp}. Here
and in the following we use the abbreviated notation $\mathbf{h}^z$,
corresponding to $\Delta\mathbf{R}=\pm a_\text{c} \hat{\mathbf{z}}$,
and $\mathbf{h}^{xz}$, corresponding to $\Delta\mathbf{R}=a_\text{c}
(\pm \hat{\mathbf{x}} \pm \hat{\mathbf{z}})$ (and analogously for all
other cartesian directions).  We note that in the A-AFM case the
translational equivalence between the two Mn sites within the unit
cell is broken, and $\Delta \mathbf{R} = \pm a_\text{c}
\hat{\mathbf{z}}$ is not a lattice vector in this case. Nevertheless,
in order to simplify the notation, we stick to the site-based index
and note that for A-AFM order a translation along $\hat{\mathbf{z}}$
is equivalent to reversing the two spin projections. In the following
we always report hopping amplitudes corresponding to hopping from and
to the Mn site at the origin, the corresponding parameters for all
other sites within the unit cell follow from symmetry. Similarly, we
do not add a spin index to the MLWF matrix elements but instead
discuss each case separately.

It can be seen that the hopping parameter between two
\mbox{$\lvert{3z^2-r^2}\rangle$}-like MLWFs along the $z$ direction,
\mbox{$h_{11}^{z}$} ($\equiv t$ in the effective model description),
is the leading term for the nearest neighbor hopping, and that overall
the next nearest neighbor hopping is about an order of magnitude
smaller than the nearest neighbor hopping. The hopping amplitude
between two \mbox{$\lvert{x^2-y^2}\rangle$}-like functions along the
$z$ direction, $h^z_{22}$ ($\equiv t'$ in the model description), is
indeed very small compared to $h^z_{11}$. In the FM case, all nearest
neighbor hopping amplitudes for the minority spin orbitals (except
$h^z_{22}$) are reduced (to about 75-85\%) compared to the majority
spin channel. This reflects the weaker hybridization between minority
spin $e_{g}$ and O(2$p$) states, leading to more localized minority
spin MLWFs with reduced hopping amplitudes.  For A-AFM order,
$h_{11}^{z}$ corresponds to the hopping between a local majority and a
local minority spin orbital, and its value, (\mbox{92\,\%} of
$h_{11}^{z}$ for FM ($\uparrow$)), is intermediate between the
corresponding FM majority and minority values. The A-AFM hopping
amplitudes within ferromagnetically ordered $x$-$y$ planes for local
majority/minority spin directions are very similar to the
corresponding FM hoppings (differing by less than \mbox{5\,meV}), with
the exception of the (local) majority spin $h_{11}^{x}$ value, which
is larger than that. Similar relations between the FM majority and
minority spin and A-AFM values are also observed for the next-nearest
neighbor hoppings.

It can easily be verified, that the hopping parameters for FM majority
and minority spin fulfill the relations described in
Eqs.~(\ref{eq:hz})-(\ref{eq:hy}), as required for cubic
symmetry. However, if the terms proportional to $t' \equiv h^z_{22}$
are neglected, the corresponding equations are not exactly
fulfilled. Thus, simply neglecting $h^z_{22}$ while keeping all other
nearest neighbor hopping amplitudes unchanged, leads to slight
deviations from cubic symmetry. Furthermore,
Eqs.~(\ref{eq:hz})-(\ref{eq:hy}) are clearly not fulfilled for the
A-AFM hopping amplitudes, which reflects the overall tetragonal
symmetry resulting from the magnetic order.

This symmetry reduction for the A-AFM case is also visible in the
on-site matrix elements $h^{0}_{11}$ and $h^0_{22}$, which differ by
about 100~meV. On the other hand, the small asymmetry ($\sim$ 1~meV)
in these on-site terms for FM order results from small numerical
accuracies during the total spread minimization (which uses the full
k-point grid, so that cubic symmetry is not automatically enforced).

Within the effective two-band model for manganites described in
Sec.~\ref{sec:lmo-model}, the Hund's rule coupling leads to an on-site
spin splitting equal to $2J$ (treating $\mathbf{S}_\mathbf{R}$ as
classical unit vector). From the calculated on-site MLWF matrix
elements, we thus obtain a value of \mbox{$J=1.499$~eV} for the Hund's
rule coupling parameter in the FM case, and \mbox{$J=1.370/1.451$~eV}
from the A-AFM on-site terms. The differences between these values
indicate the limits of the assumption of a fixed $t_{2g}$ core spin.
We note that all these values are slightly larger than the results
obtained in previous LSDA calculations
(\mbox{$J=1.34$~eV}),\cite{2007_ederer} which reflects the fact that
GGA in general leads to a stronger magnetic splitting than
LSDA.\cite{Singh/Ashkenazi:1992,Lechermann_et_al:2002}

Overall, the results obtained via MLWFs are in a very good qualitative
agreement with the previous study using TB fits to DFT band
structures.\cite{2007_ederer} However, the direct comparison between
the values calculated from MLWF in this work and the values reported
in Ref.~\onlinecite{2007_ederer} is slightly hampered by the different
exchange correlation functionals and pseudopotentials used in the two
studies. The same fitting method as described in
Ref.~\onlinecite{2007_ederer} applied to the GGA band structure
calculated in the present work, leads to a nearest neighbor hopping
parameter \mbox{$t=-688$~meV}, i.e. slightly larger than the
$-$648~meV obtained from the MLWFs. This is due to the larger majority
spin $e_g$ bandwidth obtained here, \mbox{$W_{\uparrow}=4.126$~eV},
compared to the value of \mbox{3.928~eV} reported in
Ref.~\onlinecite{2007_ederer}. Thus, the difference in bandwidth
compensates the neglect of further neighbor hopping in the simple TB
fit, leading to the apparent very good agreement between
$h^z_{11}=-648$~meV listed in Table~\ref{tab:c-hp} and the
corresponding value ($t=-655$~meV) given in
Ref.~\onlinecite{2007_ederer}.

In the following sections, we will analyze the influence of the
structural distortions only for the on-site and nearest-neighbor
hopping terms. We have verified that the resulting changes in the
further neighbor hopping amplitudes do not lead to significant
differences in the dispersion characteristics of the $e_g$ bands, even
though the corresponding relative changes of the next-nearest neighbor
hoppings are comparable with those of the nearest-neighbor hoppings.

\subsection{Jahn-Teller distortion -- structure~(ii)}\label{ss:ii}

As described in Sec.~\ref{sec:struc}, the staggered JT distortion,
$Q^x_\mathbf{R}=\pm Q^x_\mathbf{0}$, leads to a unit cell doubling
within the $x$-$y$ plane. In the case of FM order, we therefore
construct two pairs of $e_g$ MLWFs for each spin channel, localized at
the two different Mn sites within the unit cell, while for A-AFM order
we construct four pairs of MLWFs, localized at the four different Mn
sites within the corresponding unit cell (for global spin up
projection only). The same approach for choosing the energy window for
the disentanglement procedure was used as described in the previous
section.

\begin{figure}
  \includegraphics[width=\columnwidth]{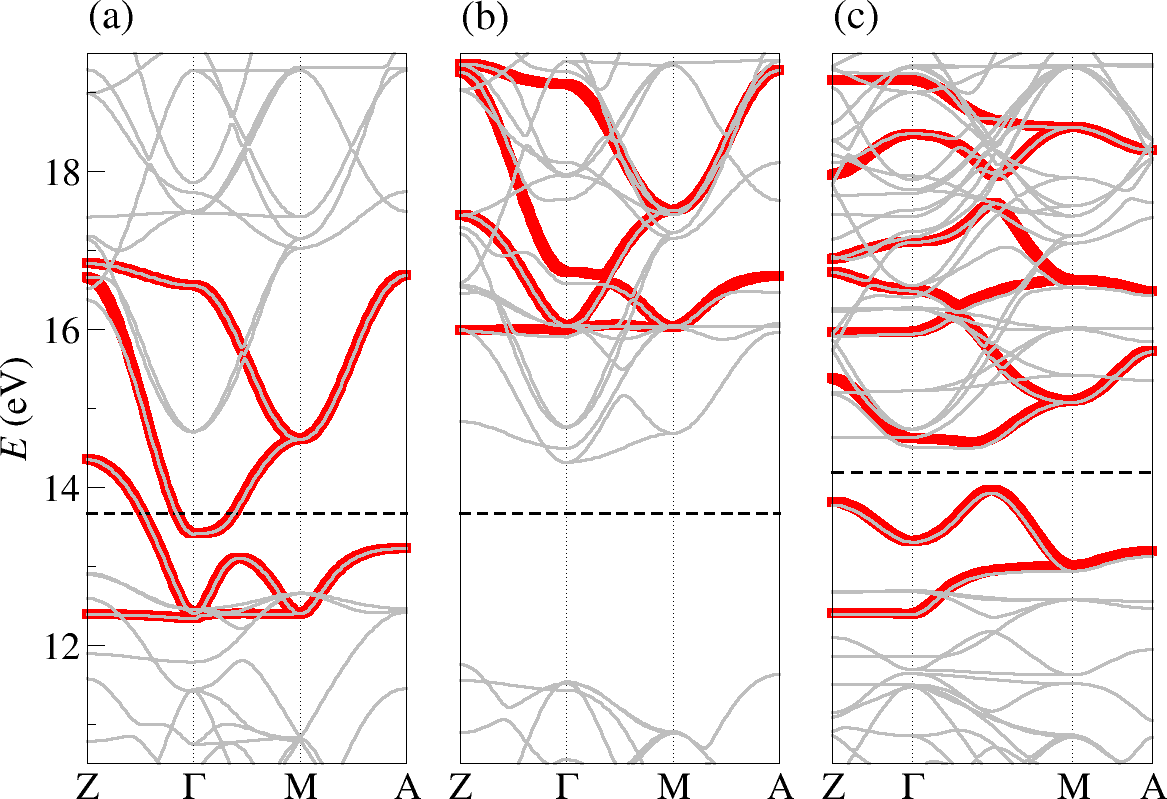}
  \caption{(Color online) DFT band structure (thin lines) for the
    JT-distorted structure~(ii): a) majority spin FM, b) minority spin
    FM, and c) A-AFM. MLWF bands are depicted as thick/red lines. The
    Fermi level is indicated by the dashed line.}
  \label{fig:jt-bs}
\end{figure}

The calculated DFT band structure and $e_{g}$-like MLWF dispersion for
the JT distorted structure~(ii) are shown in Fig.~\ref{fig:jt-bs}.  As
a result of the unit cell doubling, there are now 4 and 8 bands with
$e_{g}$ character per spin channel for the FM and A-AFM order,
respectively. As for the cubic perovskite structure, the calculated
MLWF dispersion largely follows the DFT band structure, except where
there is strong hybridization with states of a different orbital
character.  It can be seen that several degeneracies and potential
band crossings, which would result from a simple ``backfolding'' of
the cubic band-structure onto the smaller tetragonal BZ, are lifted
due to the JT distortion. This can be seen for example for the FM
majority spin bands, where the highest-lying band along
$\Gamma\mathrm{Z}$ acquires some dispersion, leading to a splitting of
the higher energy $e_g$ states at Z. Similarly, the degeneracy of the
two lowest-lying $e_g$ states at $\Gamma$ is lifted, and a potential
crossing of $e_g$ bands is prevented between $\Gamma$ and M. The
latter splitting, together with the reduced dispersion along
$\Gamma\mathrm{Z}$ for A-AFM order, appears crucial for the opening of
an energy gap in the JT-distorted A-AFM ordered structure
(Fig.~\ref{fig:jt-bs}c).

\begin{figure}
  \includegraphics[width=\columnwidth]{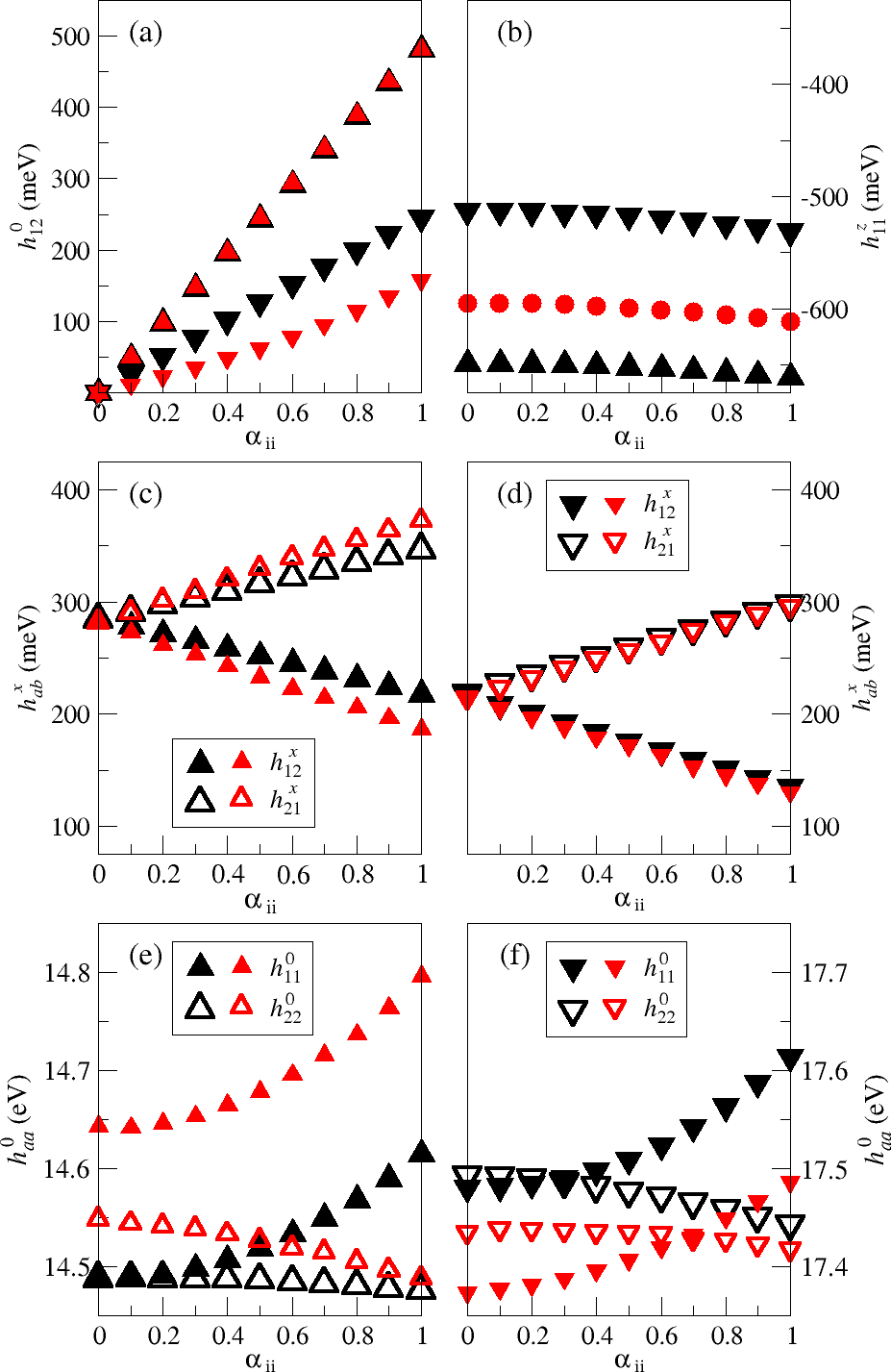}
  \caption{(Color online) MLWF Hamiltonian matrix elements
    $h^{\Delta\mathbf{R}}_{ab}$ as function of the JT distortion.
    Large/black and small/red symbols correspond to FM and A-AFM
    order, respectively. Matrix elements associated with pure (local)
    majority and minority spin character are shown as triangles
    pointing up and down, respectively. Closed circles in (b)
    represent the A-AFM $h_{11}^{z}$ hopping.}
  \label{fig:jt-h}
\end{figure}

To further analyze the influence of the JT distortion on the $e_g$
electronic structure, we perform a series of calculations where we
gradually change the oxygen positions from the ideal perovskite
structure (i) to the fully JT distorted structure (ii), according to
Eq.~\eqref{eq:vardist}, and monitor the resulting changes in the MLWF
Hamiltonian matrix elements. In all these calculations, we use the
same energy windows of $[12.0,17.5]$~eV, $[15.9,20.0]$~eV and
$[12.0,20.0]$~eV for the disentanglement in the case of FM majority,
FM minority and A-AFM, respectively. The resulting MLWF matrix
elements are depicted in Fig.~\ref{fig:jt-h}. As discussed in the
previous section, we report only hopping from and to the Mn site at
the origin. The hopping amplitudes corresponding to other sites in the
unit cell follow from symmetry. We find a strong linear dependence on
the JT distortion for both the off-diagonal on-site matrix elements
$h^0_{12}$ (Fig.~\ref{fig:jt-h}a) as well as for the off-diagonal
in-plane hopping $h^x_{12/21}$ (Fig.~\ref{fig:jt-h}c/d). All other
on-site and nearest neighbor hopping matrix elements show only a weak
or moderate quadratic dependence on $\alpha_{\mathrm{ii}}$.

Within the model described in Sec.~\ref{sec:lmo-model} the sole effect
of the JT distortion $(Q^x_\mathbf{R},Q^z_\mathbf{R})$ is a linear
coupling to the on-site terms at site $\mathbf{R}$ according to:
\begin{equation}
\label{eq:JT-local}
\mathbf{t}^0 = \left(
\begin{array}{cc}
e_0 - \lambda Q^z_\mathbf{R} & - \lambda Q^x_\mathbf{R} \\ - \lambda
Q^x_\mathbf{R} & e_0 + \lambda Q^z_\mathbf{R} \end{array} \right) \ .
\end{equation}
In our case $Q^z_\mathbf{R} = 0$ and $Q^x_\mathbf{R} = \pm
\alpha_{\mathrm{ii}} Q^x_\mathbf{0}$; $e_0$ is the on-site energy of
the $e_g$ orbitals. It can be seen from Fig.~\ref{fig:jt-h}a that the
off-diagonal element $h^0_{12}$ indeed shows a linear dependence on
$\alpha$, consistent with Eq.~(\ref{eq:JT-local}). The corresponding
slope, $-\lambda Q^x_\mathbf{0}=482$~meV, is nearly identical for the
FM majority and A-AFM local majority spin elements, whereas it is
significantly smaller for the (local) minority spin matrix elements
($-\lambda Q^x_\mathbf{0}=246/155$~meV). This indicates that the JT
splitting is also a ligand-field effect, i.e. it is mediated by
hybridization with the surrounding oxygen orbitals, which, as pointed
out previously, is stronger for the energetically lower majority spin
states. The values for the JT coupling constant $\lambda$ obtained
from the data shown in Fig.~\ref{fig:jt-h}a are \mbox{3.19~eV/\AA},
\mbox{1.63~eV/\AA}, and \mbox{1.02~eV/\AA}, for majority, FM minority,
and A-AFM local minority spin states, respectively. We note that the
value of $\lambda$ obtained for majority spin is approximately a
factor of two larger than the value obtained from the fitting
procedure described in Ref.~\onlinecite{2007_ederer}. As we will
discuss in more detail below, the source for this discrepancy is the
strong linear splitting observed for the off-diagonal in-plane nearest
neighbor hoppings $h^x_{12/21}$, which is induced by the JT distortion
(see Fig.~\ref{fig:jt-h}c/d).

This splitting between $h^x_{12/21}$ again results from the underlying
hopping between atomic Mn($e_g$) and O($p$) states, which (in leading
order) depends linearly on the \mbox{Mn-O} distance. Since this
dependence will be different for the $|3z^2-r^2\rangle$ and
$|x^2-y^2\rangle$ orbitals, it can easily be verified that the
effective hopping across a combination of one long and one short Mn-O
bond within the $x$-$y$ plane between two different $e_g$ orbitals
will also depend linearly on the JT distortion, whereas the effective
hopping between the same type of $e_g$ orbitals will show only a
quadratic dependence. We have verified, by constructing atomic-like
Wannier functions for both Mn($e_g$) and O($p$) orbitals
(corresponding to larger energy windows), that indeed the dependence
on the \mbox{Mn-O} distance is much stronger for the hopping amplitude
between the $|3z^2-r^2\rangle$-type orbital and a neighboring O($p$)
orbital than for the corresponding $|x^2-y^2\rangle$-type hopping,
consistent with the observed splitting in the effective hopping
amplitudes $h^x_{12/21}$ shown in Fig.~\ref{fig:jt-h}c/d.

It can be verified within a TB model where the linear splitting
between $h^x_{12}$ and $h^x_{21}$ (and analogously for the hopping
along the $y$ direction) is taken into account via one extra parameter
derived from the MLWF data, that this splitting partially cancels the
effect of the on-site JT term on the band dispersion. In particular,
the JT-induced ``gap'' between the second and third $e_g$ band between
$\Gamma$ and M is reduced by increasing the $h_{12}^x$/$h_{21}^x$
splitting, whereas it is enhanced by increasing the JT coupling
strength $\lambda$. Thus, the band dispersion resulting from reduced
$\lambda$ and no splitting between $h_{12}^x$ and $h_{21}^x$ looks
very similar to the one obtained from the MLWF parameters
(i.e. including the spitting between $h^x_{12/21}$). This is the
reason why the fitting of the DFT band structure on a TB model that
does not incorporate a $h_{12}^x$/$h_{21}^x$ splitting (see
Ref.~\onlinecite{2007_ederer}) leads to a smaller value of $\lambda$
than the one obtained from the MLWF parameters. An interesting
question arising from this is whether, despite the very similar band
dispersion, the two different TB parameterizations would lead to
noticeable differences in calculated ordering temperatures for the
collective JT distortion.

The differences between the off-diagonal in-plane hopping parameters
induced by the JT distortion indicate changes of the MLWFs themselves,
i.e. the JT distortion alters the basis-set of a MLWF-based TB model.
We note that this is an unavoidable result of the effective
``two-band'' picture. The definition of a distortion-independent
basis-set is only possible within a full $d$-$p$ TB model, based on
truly atomic-like functions. On the other hand, a splitting between
$h_{12}^x$ and $h_{21}^x$ can in principle also result from a unitary
mixing of the $|3z^2-r^2\rangle$ and $|x^2-y^2\rangle$ basis
functions. In order to check whether (at least part of) the observed
splitting is due to such a mixing, we have applied a local unitary
transformation between the two MLWFs on each site, and studied the
resulting changes in the MLWF matrix elements.  In essence, we find
that it is impossible to retrieve the ``cubic symmetry'', i.e.  the
form described in Eqs.~(\ref{eq:hz})-(\ref{eq:hy}) and
(\ref{eq:JT-local}), simultaneously for $\mathbf{h}^0$,
$\mathbf{h}^z$, and $\mathbf{h}^x$, and that a transformation of one
of these terms to the desired form in general increases the
corresponding deviations in the other two terms.  It appears that the
basis functions resulting directly from the maximum localization
procedure using initial projections on atomic $|3z^2-r^2\rangle$ and
$|x^2-y^2\rangle$ functions represent the best overall compromise.

The leading hopping term in $z$ direction, $h_{11}^{z}$
(Fig.~\ref{fig:jt-h}b), exhibits only a weak quadratic change as a
function of $\alpha_\mathrm{ii}$. We also find a similar weak
quadratic dependence on the JT distortion in the hopping parameters
$h_{11/22}^{x}$ (not shown), and a moderately strong quadratic change
in the on-site diagonal matrix elements (Fig.~\ref{fig:jt-h}e/f),
which introduces a splitting of about \mbox{150~meV} between
$h_{11}^{0}$ and $h_{22}^{0}$ for the fully JT distorted structure.

Finally, we note that the Hund's rule coupling parameters derived from
the local spin splitting between MLWFs obtained for the fully JT
distorted structure (\mbox{$J=1.499/1.484$~eV} for FM order,
\mbox{$J=1.345/1.465$~eV} for A-AFM order) are not significantly
changed compared with the ones obtained for structure (i).

\subsection{GdFeO$_{3}$-type distortion --
  structure~(iii)}\label{ss:iii}

\begin{figure}
  \includegraphics[width=\columnwidth]{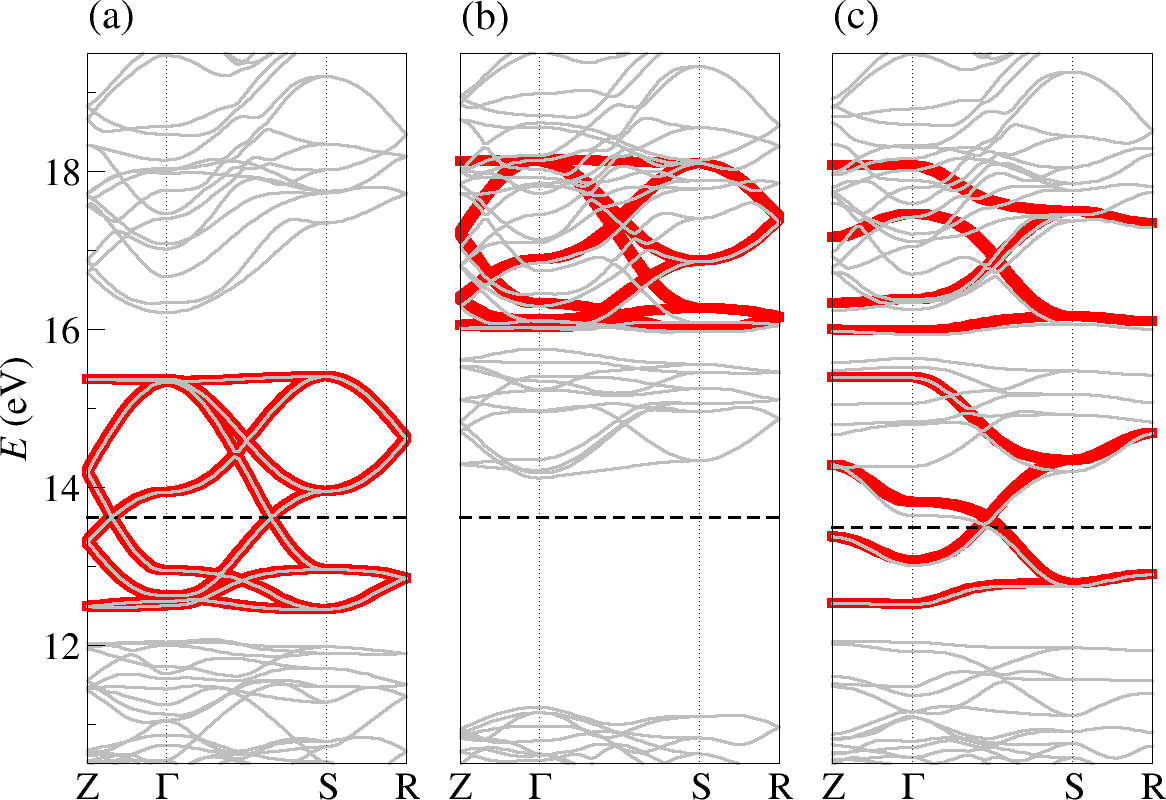}
  \caption{(Color online) DFT band structure (thin lines) for the
    purely GFO-distorted structure~(iii): a) FM majority spin, b) FM
    minority spin, c) A-AFM.  MLWF bands are depicted as thick/red
    lines. The Fermi level is indicated by dashed lines.}
  \label{fig:gfo-bs}
\end{figure}

The band dispersion calculated for the purely GFO-distorted structure
(iii) is presented in Fig.~\ref{fig:gfo-bs}.  The rotation and tilting
of oxygen octahedra in structure (iii) distorts the ideal 180$^\circ$
\mbox{Mn-O-Mn} bond angle, which is expected to reduce the hopping
amplitudes. Indeed, it can be seen in Fig.~\ref{fig:gfo-bs} that the
GFO distortion leads to significantly smaller bandwidth
(\mbox{2.951~eV} and \mbox{2.139~eV} for FM majority and minority
spin, respectively, compared to \mbox{4.126~eV} and \mbox{3.156~eV} in
the undistorted structure (i)). As a result, the FM majority spin
$e_{g}$ bands become completely separated from the lower-lying
$t_{2g}$ bands and the La($d$) bands at higher energy. Unlike in the
JT distorted structure (ii), the system stays metallic for both FM and
A-AFM order.

Since the unit cell for structure (iii) is quadrupled with respect to
the cubic perovskite structure, there are now 8 bands with dominant
$e_{g}$ character for each spin direction. However, due to the
tilt/rotation of the oxygen octahedra, ``$e_g$-like'' orbitals at a
certain site can hybridize with ``$t_{2g}$-like'' orbitals at a
neighboring site, leading to bands with mixed $e_{g}$/$t_{2g}$
character.\cite{footnote:eg-t2g} In the FM case this does not
represent a problem for the disentanglement procedure, since the bands
with predominant $e_g$ character are separated from the predominantly
$t_{2g}$ bands for both spin direction. For FM order, we can therefore
construct four pairs of MLWFs, localized at the four different sites
within the unit cell, by defining appropriate energy windows
separately for each spin direction. This is not possible in the A-AFM
case, where the local minority $t_{2g}$ bands overlap strongly with
the local majority $e_g$ bands in the energy region between
\mbox{14~eV} and \mbox{16~eV}. In this case, the standard
disentanglement procedure employed for structures (i) and (ii),
i.e. defining an energy window \mbox{$[12.0,21.0]$~eV} and
initializing 8 Wannier functions from projections on atomic $e_g$
orbitals at the various sites, results in MLWFs with mixed
$t_{2g}$/$e_g$ orbital character. In particular, the resulting local
minority spin MLWFs exhibit a rather strong $t_{2g}$ character.

One possible way to overcome this problem would be to construct all 20
$d$-like MLWFs (5 per Mn site), i.e. both $e_g$ and $t_{2g}$
orbitals. However, the resulting MLWFs still contain some amount of
$e_g$/$t_{2g}$ mixing, and the corresponding MLWF matrix elements
exhibit systematic deviations from the results obtained in the
previous sections, which are derived from a smaller set of MLWFs. In
the following, we therefore adopt a different strategy to obtain model
parameters for the A-AFM case, and construct the 4 local majority and
4 local minority spin $e_{g}$-like MLWFs separately, using two
different energy windows. From this, we obtain the on-site matrix
elements $\mathbf{h}^{0}$ as well as the hopping parameters
$\mathbf{h}^{x}$ within the $x$-$y$ plane (and of course all further
neighbor hopping amplitudes within this plane). On the other hand we
do not obtain the hopping amplitudes $\mathbf{h}^{z}$ between adjacent
planes in the $z$ direction, which would connect the two separate sets
of MLWFs. Similar to the purely JT distorted case, we analyze the
effect of the GFO distortion on the $e_g$ bands by performing
calculations with varying degree of distortion, i.e. by changing the
oxygen positions according to Eq.~\eqref{eq:vardist}. In this case we
always adjust the energy window for the construction of the MLWFs to
the actual $e_{g}$ bandwidth corresponding to a particular
$\alpha_{\mathrm{iii}}$.

\begin{figure}
  \includegraphics[width=\columnwidth]{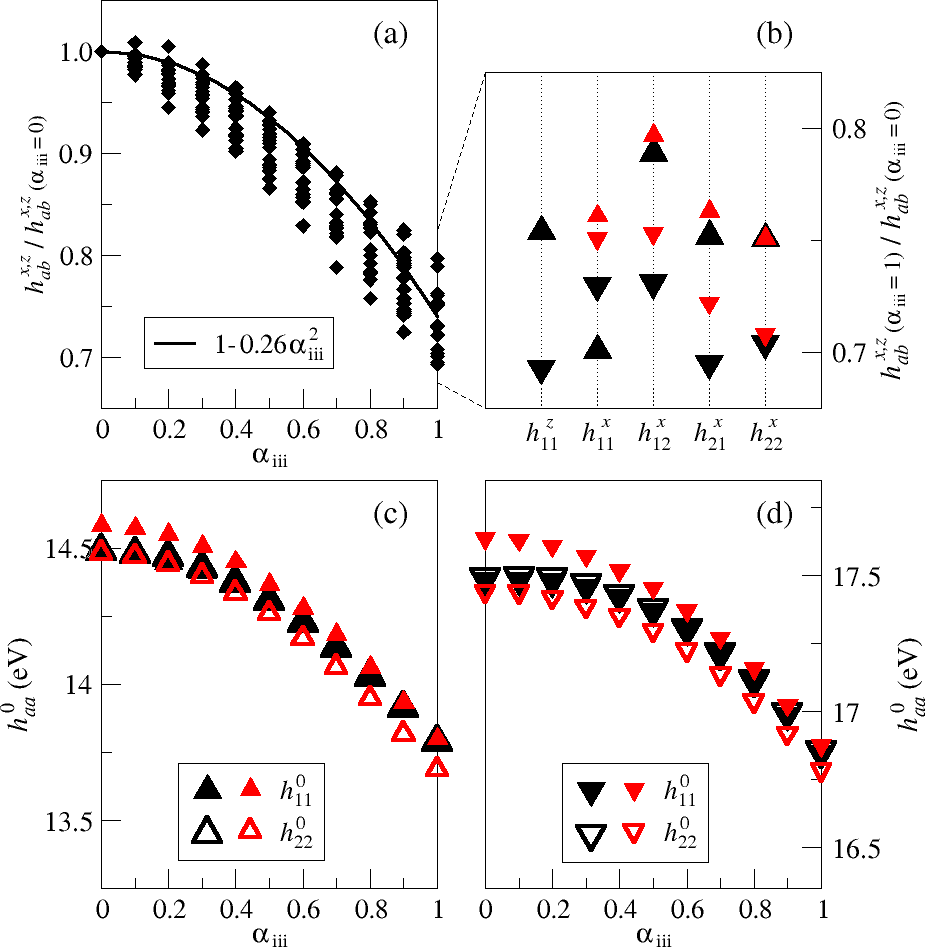}
  \caption{(Color online) Hamiltonian matrix elements in the basis of
    MLWFs as a function of the GFO distortion. Large/black and
    small/red symbols correspond to FM and A-AFM order, respectively.
    Elements associated with purely (local) majority and minority spin
    character are represented by triangles pointing up and down,
    respectively.}
  \label{fig:gfo-h}
\end{figure}

We find that the main effect of the GFO distortion is indeed a
systematic reduction of all hopping amplitudes by $\approx 20-30$\,\%,
consistent with what was reported in
Ref.~\onlinecite{2007_ederer}. Fig.~\ref{fig:gfo-h}a shows the overall
reduction for all obtained nearest neighbor hopping amplitudes for
both FM and A-AFM order, while Fig.~\ref{fig:gfo-h}b resolves the
reduction factors of the various hopping amplitudes for full GFO
distortion ($\alpha_\text{iii}=1$). It can be seen, that even though
there is a significant spread in the reduction factors for the various
hopping parameters, the overall reduction can approximately be
described as $\mathbf{h}^{x/z}(\alpha_\mathrm{iii}) =
\mathbf{h}^{x/z}(0) \left({1-\eta\alpha^{2}_{\mathrm{iii}}}\right)$,
with an average value of $\eta=0.26$.

In addition to the changes in the nearest neighbor hopping amplitudes,
we also observe a quadratic decrease of the on-site diagonal matrix
elements as a function of the GFO distortion
(Fig.~\ref{fig:gfo-h}c/d), with a similar magnitude for both orbitals
and different magnetic order. This can be understood again from the
underlying hopping between atomic $p$ and $d$ orbitals. Since the
effective $e_g$ bands correspond to the antibonding combination of
these atomic orbitals, a reduction of the underlying $p$-$d$ hopping
amplitudes results in a decrease of the $\Gamma$-point energy of the
$e_{g}$ states. The Hund's rule coupling parameter \mbox{$J=1.502$~eV}
obtained from the on-site splitting for FM order and
$\alpha_\mathrm{iii}=1$ is very similar to the corresponding value for
the cubic perovskite structure.

\subsection{Combined Jahn-Teller and GdFeO$_{3}$-type distortion --
  structure~(iv)}\label{ss:iv-v}

\begin{figure}
  \includegraphics[width=\columnwidth]{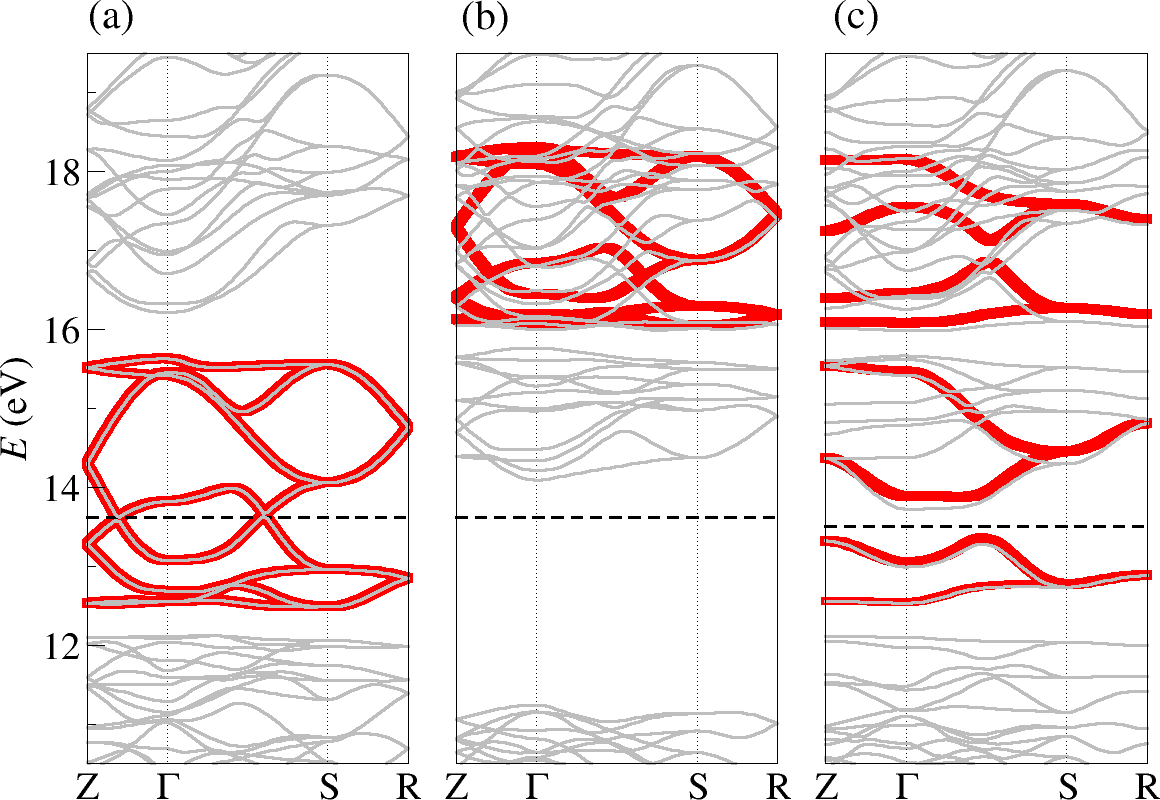}
  \caption{(Color online) DFT band structure (thin lines) for
    structure~(iv): a) majority spin FM, b) minority spin FM, and c)
    A-AFM.  MLWF are depicted as thick/red lines. Fermi level is
    indicated by dashed line.}
  \label{fig:jtgfo-bs}
\end{figure}

So far we have analyzed the individual effects of the JT and GFO
distortion. We now discuss whether the superposition of both
distortions gives rise to any changes in the MLWF matrix elements that
go beyond a simple superposition of the individual effects. The
corresponding band structure and MLWF dispersion for structure (iv),
i.e. the combined JT and GFO distortion, is presented in
Fig.~\ref{fig:jtgfo-bs}.  It can be seen that the band structure in
this case closely resembles the one of the purely GFO distorted
structure~(iii), Fig.~\ref{fig:gfo-bs}, but with the additional
JT-induced effects (avoided band-crossings and lifted degeneracies) as
described in Sec.~\ref{ss:ii}. Note that, as in the purely JT
distorted structure, the FM case is metallic, whereas a band gap opens
only for A-AFM order.

\begin{figure}
  \includegraphics[width=\columnwidth]{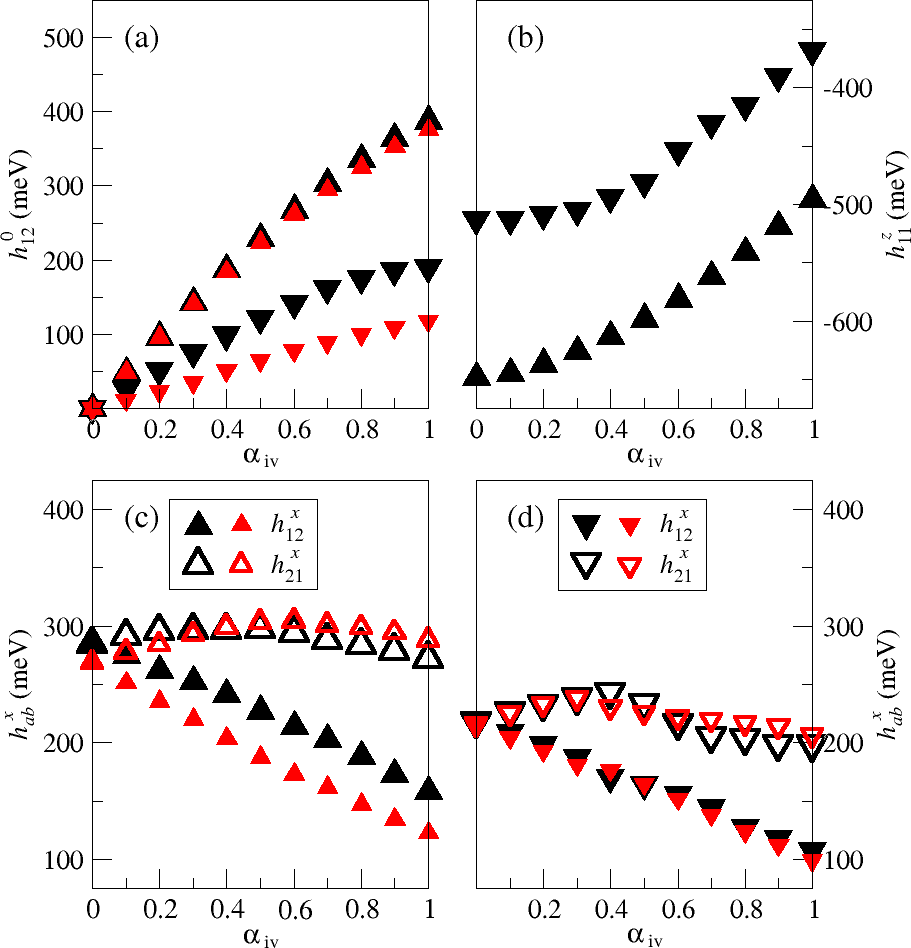}
  \caption{(Color online) MLWF Hamiltonian matrix elements as function
    of combined JT and GFO distortion. Large/black and small/red
    symbols correspond to the FM and A-AFM order, respectively.
    Elements associated with purely (local) majority and minority spin
    character are represented by triangles pointing up and down,
    respectively.}
  \label{fig:jtgfo-h}
\end{figure}

As described in the previous section we construct 8 MLWFs per spin
direction for the FM case and two separate sets of 4 local majority
and 4 local minority $e_{g}$-like MLWFs for the A-AFM
case. Fig.~\ref{fig:jtgfo-h} shows the evolution of selected MLWF
matrix elements as a function of distortion. The atomic positions are
changed according to Eq.~\eqref{eq:vardist} with
\mbox{$x=\mathrm{iv}$}.  By comparing Fig.~\ref{fig:jtgfo-h}a with
Fig.~\ref{fig:jt-h}a, it can be seen that the GFO distortion does also
significantly reduce the on-site matrix elements $h_{12}^{0}$ (to
\mbox{$\approx 75-80\,\%$}), which are otherwise proportional to the
JT distortion. This is further evidence for the ligand-field nature of
the JT coupling, i.e. that it is mediated by the Mn-O hybridization
(which is reduced by the GFO distortion).  Furthermore, it can be seen
that the leading hopping along $z$, $h_{11}^{z}$, follows very closely
the trend observed for the purely GFO distorted structures. In the
case of the off-diagonal hopping amplitudes within the $x$-$y$ plane,
the superposition of GFO-distortion-induced reduction and JT-induced
splitting leads to an initial increase of $h_{12}^x$ for small
distortion, followed by a decrease for larger
$\alpha_{\mathrm{iv}}$. Overall, the observed trends can indeed be
well understood as independent superposition of the individual effects
of JT and GFO distortions. We note that the kinks observed in some of
the minority spin hopping terms around $\alpha_{\mathrm{iv}}\approx
0.4$ result from the opening of the gap between $e_{g}$-like and
$t_{2g}$-like minority spin bands for this amount of distortion, which
represents a certain ``discontinuity'' for the disentanglement
procedure.

\subsection{Simplified TB models for LaMnO$_3$ in the full experimental
$Pbnm$ structure (v)} \label{ss:tb}

The analysis presented so far showed that the effect of different
structural distortions on the $e_g$ bands can, to a good extent, be
treated independently of each other. In this section, we attempt to
incorporate the most significant effects described in the previous
sections into a refined effective TB model. Then, in order to test the
accuracy of the resulting parameterization, we compare the resulting
band dispersion with the full GGA and MLWF band-structure, calculated
for the full experimental $Pbnm$ structure of LaMnO$_3$ and A-AFM
order.

For the refined TB model we introduce different hopping amplitudes for
local majority/minority spin projections to describe the hopping
between ferromagnetically aligned nearest neighbors within the $x$-$y$
planes ($t^{\uparrow\uparrow}$/$t^{\downarrow\downarrow}$), and an
intermediate value for the nearest neighbor hopping between
antiferromagnetically aligned nearest neighbors along the $z$
direction ($t^{\uparrow\downarrow}$), i.e. hopping between two
different local spin projections.  This is in accordance with our
results presented in Sec.~\ref{ss:i}. For the corresponding hopping
amplitudes we use the values of $h_{11}^{z}$ calculated for the ideal
cubic perovskite structure (see Table~\ref{tab:c-hp}) for FM and A-AFM
order, which are then reduced by the same factor $(1-\eta
\alpha_{\mathrm{iii}}^2)$, where $\alpha_{\mathrm{iii}}$ describes the
amount of pure GFO distortion. Apart from these modifications we
assume the usual cubic symmetry of the nearest neighbor hopping
matrices, i.e.:
\begin{align}
\mathbf{t}^{ss'}(\pm a_\text{c} \hat{\mathbf{z}}) & = (1 - \eta
\alpha_{\mathrm{iii}}^2) \, t^{ss'} \left(
    \begin{matrix}
      1 & 0 \\ 0 & 0
    \end{matrix} \right) \ ,
\\ \mathbf{t}^{ss'}(\pm a_\text{c} \hat{\mathbf{x}}) & = (1 - \eta
\alpha_{\mathrm{iii}}^2)\, t^{ss'} \left(
    \begin{matrix}
      \tfrac14 & -\tfrac{\sqrt{3}}{4}\\ -\tfrac{\sqrt{3}}{4} &
      \tfrac34
    \end{matrix} \right)
\ ,
\end{align}
(and analogously for $\mathbf{t}^{ss'}(\pm a_\text{c}
\hat{\mathbf{y}})$). Note that $s$ and $s'$ in these equations should
be read as a \emph{local} spin index, i.e. it designates the spin
projection relative to the orientation of the local core spin. We use
the average value $\eta=0.26$ determined in Sec.~\ref{ss:iii}.

The JT-induced splitting of the non-diagonal elements of the hopping
matrix within the $x$-$y$ plane discussed in Sec.~\ref{ss:ii} is
incorporated in the TB model as an additional contribution to the
in-plane hopping:
\begin{equation}
  \Delta \mathbf{t}(\pm a_\text{c} \hat{\mathbf{x}}) = \tilde{\lambda}
  Q^x_\mathbf{0} \alpha_{\mathrm{ii}}(1-\eta \alpha_{\mathrm{iii}})
  \left( \begin{matrix} 0 & 1 \\ -1 & 0
  \end{matrix} \right) 
\end{equation}
(and analogously for $\Delta \mathbf{t}(\pm a_\text{c}
\hat{\mathbf{y}})$). Here, $\alpha_{\mathrm{ii}}$ describes the
amplitude of the staggered JT distortion, i.e.
\mbox{$Q^x_\mathbf{R}=\pm Q^x_\mathbf{0} \alpha_{\mathrm{ii}}$}, and
the parameter $\tilde{\lambda}$ is determined from the average
splitting over all hopping amplitudes in the purely JT-distorted
structure (shown in Fig.~\ref{fig:jt-h}c/d). In addition, we include
the usual on-site JT effect in essentially the same form as described
in Eq.~(\ref{eq:jt}), but with a spin-dependent JT coupling constant
that is also reduced by the GFO distortion (with the same factor as
the hopping amplitudes):
\begin{equation}
\lambda \rightarrow \lambda^s (1-\eta \alpha_{\mathrm{iii}}^2) \ .
\end{equation}
We note that the orthorhombic strain in the experimental structure of
LaMnO$_3$ gives rise to a homogeneous $Q^z$ component to the JT
distortion, i.e. the same $Q^z_\mathbf{R} \neq 0$ on all sites, which
we take into account within the model by using the same coupling
constant $\lambda^s$ as for the $Q^x$ component.

We also include hopping between next nearest neighbors and between
second nearest neighbors along the cartesian coordinate axes in the
refined TB model, but we do not consider any spin-dependence of the
corresponding hopping amplitudes. We describe the hopping between
next-nearest neighbors by spin-independent parameters $t^{xy}$
corresponding to the hopping between two
$\lvert{3z^2-r^2}\rangle$-type orbitals along the $\pm a_\text{c}
\hat{\mathbf{x}} \pm a_\text{c} \hat{\mathbf{y}}$ directions. The
parameter $t^{xy}$ is taken as spin average over the corresponding
MLWF matrix elements $h^{xy}_{11}$ calculated for the cubic
structure. All other hopping matrix elements between next nearest
neighbors are determined from this via the following relations, which
are derived assuming cubic symmetry and indirect hopping only (see
Ref.~\onlinecite{2007_ederer}):
\begin{align}
\label{eq:t2xz-par}
\mathbf{t}^{xz}&= t^{xy}
\left({1-\eta\alpha_{\mathrm{iii}}^{2}}\right)
  \begin{pmatrix}
    -2&\sqrt{3}\\
    \sqrt{3}&0
  \end{pmatrix}
\\
\label{eq:t2xy-par}
\mathbf{t}^{xy}&=
t^{xy}
\left({1-\eta\alpha_{\mathrm{iii}}^{2}}\right)
  \begin{pmatrix}
    1&0\\
    0&-3
  \end{pmatrix}
\end{align}
The same GFO-distortion-induced reduction as for the nearest neighbor
hopping matrices is applied. The hopping between second nearest
neighbors along the coordinate axes [$\mathbf{t}(\pm 2a_\text{c}
\hat{\mathbf{x}})$, $\mathbf{t}(\pm 2a_\text{c} \hat{\mathbf{y}})$,
$\mathbf{t}(\pm 2a_\text{c} \hat{\mathbf{z}})$] is included according
to the ideal cubic symmetry relations described by
Eqs.~(\ref{eq:hz})-(\ref{eq:hy}), with $a_\text{c}$ replaced by
$2a_\text{c}$, $t'=0$, and $t=t^{2z}$, where $t^{2z}$ is estimated
from the MLWF matrix elements for the purely GFO distorted structure.
We note that the reduction of this parameter compared to the
undistorted case is significantly stronger than for the nearest (and
next nearest) neighbor hopping amplitudes. Furthermore, the hopping
between third nearest neighbors along the coordinate axes
[$\mathbf{t}(\pm 3a_\text{c} \hat{\mathbf{x}})$, $\mathbf{t}(\pm
3a_\text{c} \hat{\mathbf{y}})$, $\mathbf{t}(\pm 3a_\text{c}
\hat{\mathbf{z}})$], that was considered in Sec.~\ref{ss:i}, becomes
negligible as result of the GFO distortion.

Finally, we include the Hund's rule coupling in the refined TB model
using the standard form [Eq.~\eqref{eq:hund}] with an average value of
$J$ obtained from the MLWF on-site splitting. In order to relate the
obtained TB bands to the full GGA and MLWF band-structures, we
determine the on-site energy $e_{0}$ as the spin and orbital average
of the corresponding $h_{aa}^{0}$ matrix elements for the A-AFM
experimental structure.

The values of all parameters used in the refined TB model are
summarized in Table~\ref{tab:tbparam}. The JT distortion in the
experimental $Pbnm$ structure corresponds to $\alpha_\mathrm{ii}=1$,
\mbox{$Q^{x}_{\mathbf{0}}=-0.161$~\AA}, and \mbox{$Q^{z}=-0.048$~\AA},
and the corresponding amplitude of the GFO distortion is
$\alpha_{\mathrm{iii}}=1$.

\begin{table}
  \caption{Parameters used in the TB models.}
  \label{tab:tbparam}
  \begin{ruledtabular}
    \begin{tabular}{lcc}
      & refined & simple \\
      \hline\\[-8pt]
      $t^{\uparrow\uparrow}$ (eV)         & -0.648  & -0.492 \\
      $t^{\downarrow\downarrow}$ (eV)      & -0.512  & -0.492 \\
      $t^{\uparrow\downarrow}$ (eV)       & -0.569  & -0.492 \\
      $\eta$                              & 0.26    & $-$  \\
      $\tilde{\lambda}$ (eV\AA$^{-1}$)    & 0.53    & 0    \\
      $\lambda^\uparrow$ (eV\AA$^{-1}$)   & 3.19    & 1.64 \\
      $\lambda^\downarrow$ (eV\AA$^{-1}$) & 1.33    & 1.64 \\
      $t^{xy}$ (eV)                       & -0.018  & 0 \\
      $t^{2z}$ (eV)                       & -0.020  & 0 \\
      $e_{0}$ (eV)                        & 15.356  & 15.505 \\
      $J$ (eV)                            & 1.5     & 1.805 \\
    \end{tabular}
  \end{ruledtabular}
\end{table}

We also compare with a very simple TB model that includes only nearest
neighbor hopping according to Eqs.~(\ref{eq:hz})-(\ref{eq:hy}) with
$t'=0$, and the standard JT and Hund's rule coupling as described by
Eqs.~(\ref{eq:jt}) and (\ref{eq:hund}). The parameters for this model
are chosen via typical simplified fitting procedures: the nearest
neighbor hopping parameter $-t$ is obtained as one sixth of the
majority spin bandwidth $W$ for the fully GFO distorted
structure~(iii) and FM order; the JT coupling constant $\lambda$ is
taken from Ref.~\onlinecite{2007_ederer}, where it was obtained by
fitting a similar TB model (including also next nearest neighbor
hopping) to a DFT band-structure; $J$ is calculated from the spin
splitting between FM majority and minority bands at the $\Gamma$-point
for the cubic structure~(i); and $e_0$ is fitted such that the Fermi
energy is aligned with the DFT calculation value.

\begin{figure}
  \includegraphics[width=\columnwidth]{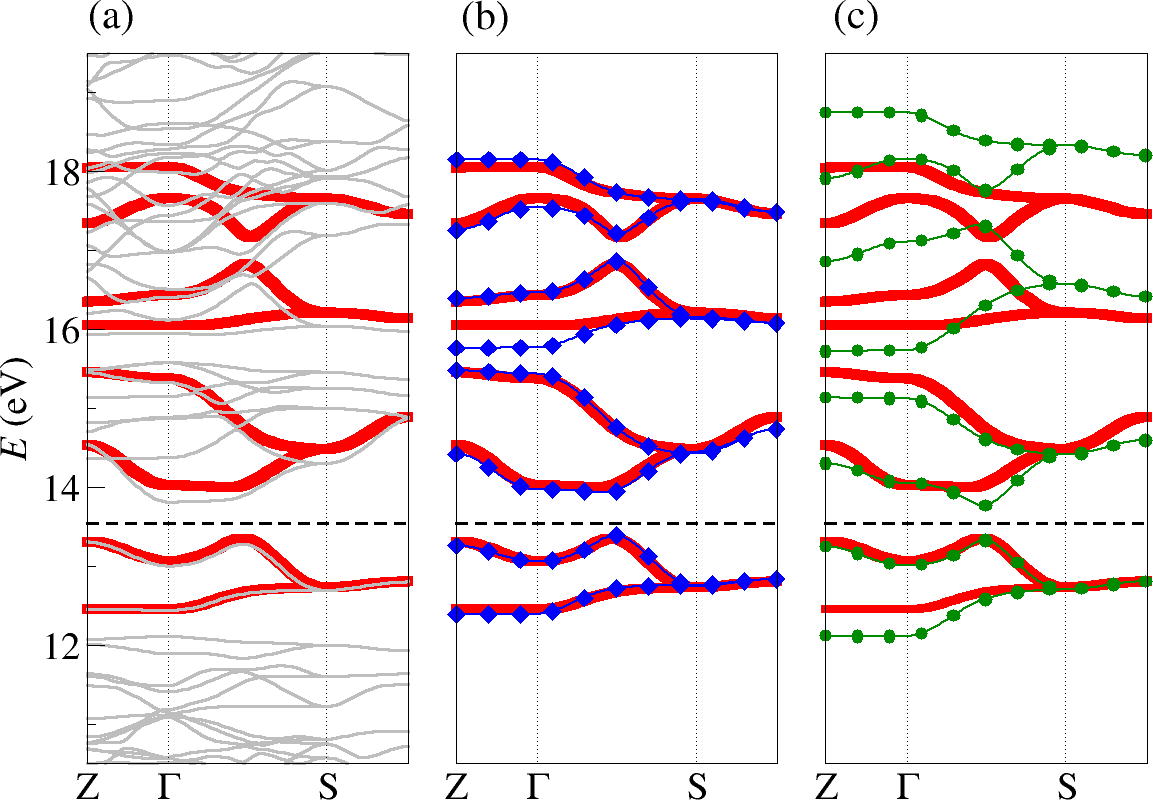}
  \caption{(Color online) (a) DFT bands (thin/grey lines) and MLWF
    bands (thick/red lines) for the A-AFM experimental $Pbnm$
    structure~(v). Comparison of the MLWF bands with the refined TB
    model (b) and with the simple nearest neighbor TB model (c) The
    Fermi level is indicated by dashed lines.}
  \label{fig:pbnm-bs}
\end{figure}

Figure~\ref{fig:pbnm-bs}a shows the band dispersion obtained from the
GGA calculation for the full experimental structure~(v) and A-AFM
order as well as the corresponding MLWF
bands. Fig.~\ref{fig:pbnm-bs}b/c shows the comparison between the MLWF
bands and the two different simplified TB models.  It can be seen that
the orthorhombic lattice strain and La displacements do not lead to
significant qualitative changes in the band-structure as compared to
structure (iv) (see Fig.~\ref{fig:jtgfo-bs}c). The comparison between
the MLWF dispersion and the refined TB model (Fig.~\ref{fig:pbnm-bs}b)
shows that, despite the many simplifications made, this model
reproduces the MLWF bands to a remarkable accuracy. The only major
discrepancy can be seen for the lowest-lying local minority band along
$\Gamma$-Z at $E \sim 16$~eV, which is slightly lower than the
corresponding MLWF band. This can be traced back to an overestimation
of the $h_{22}^{x}$ hopping amplitude, which results from the fact
that we use the same reduction factor $\eta$ for all hoppings. As can
be seen in Fig.~\ref{fig:gfo-h}b, $h^x_{22}$ is affected more strongly
by the GFO distortion than any other nearest neighbor hopping (for
A-AFM order). The very simple nearest neighbor TB model depicted in
Fig.~\ref{fig:pbnm-bs}c deviates much stronger from the MLWF band
structure than the refined model, but still captures the overall
dispersion surprisingly well. Consistent with our analysis from the
previous sections, the deviations are more pronounced for the
energetically higher local minority spin bands, which is clearly due
to the neglected spin dependence of the hopping. As discussed in
Sec.~\ref{ss:ii}, the smaller JT coupling constant used in the simple
model partially cancels the missing effect of the JT distortion on the
inter-orbital in-plane hopping parameters, leading to the relative
good agreement of the simpler model with the MLWF bands around the
Fermi level.

\section{Summary and Conclusions}
\label{sec:summary}

We have shown that the construction of maximally localized Wannier
functions together with the disentanglement procedure described in
Ref.~\onlinecite{2001_souza} can be used to extract effective $e_g$
bands in LaMnO$_3$ even for cases where these bands are strongly
entangled with other states. This procedure thus provides a very
robust way for extracting the ``correlated subspace'' used for example
in DFT+DMFT calculations.

We have used this procedure to obtain a TB parameterization of the
$e_g$ bands for different structural modifications of LaMnO$_3$ with
both FM and A-AFM order. By monitoring the effect of the individual
distortions on the MLWF matrix elements, we can assess the quality of
the various approximations and simplifications that are commonly used
in model Hamiltonians for manganite systems. In particular, we find
the following:
\begin{itemize}
\item{While the nearest neighbor hopping is clearly dominant, the
  further neighbor hopping along the cartesian axes decays rather
  slowly. Taking into account nearest, next-nearest, as well as second
  and third nearest hopping along the cartesian axes leads to
  deviations of less than 0.11~eV from the (cubic FM) DFT band
  structure.}
\item{In addition to the linear on-site coupling to the JT distortion,
  we observe a strong effect on the in-plane hopping amplitudes
  between different orbitals. The corresponding splitting, which is
  due to the underlying Mn-O hopping, partially cancels the effect of
  the on-site term on the band dispersion, which has a strong
  influence on the determination of the local JT coupling strength.}
\item{The GFO distortion leads to an overall reduction of all hopping
  amplitudes by about 25-30~\%, and also reduces the local JT
  splitting. This reduction is due to the weaker hybridization between
  Mn($e_g$) and O($p$) states for non-180$^\circ$ bond angle.}
\item{The higher energy of the (local) minority spin states reduces
  the hybridization between the corresponding atomic $e_g$ and O($p$)
  states, leading to reduced hopping amplitudes and JT coupling
  compared to the majority spin states.}
\item{The splitting between (local) majority and minority spin states
  is generally well described by the local Hund's rule coupling, even
  though small variations in the corresponding $J$ values indicate the
  limits of the core spin approximation.}
\end{itemize}

It is apparent that the most crucial deviations from the simple two
band description are a result of the underlying Mn-O hybridization.
Nevertheless, we have shown that a refined TB model that incorporates
the effects described above using the parameters listed in
Table~\ref{tab:tbparam} reproduces the DFT band structure calculated
for the full experimental crystal structure of LaMnO$_3$ with
remarkable accuracy. Whether this accuracy, at the prize of more
parameters in the model, is desirable depends of course on the
specific application of the model description.
 
Furthermore, our analysis shows that the effects of the various
distinct structural distortions present in LaMnO$_3$ are (to a good
approximation) independent from each other and can therefore be
assessed individually. However, the GFO distortion has to be taken
into account to obtain the correct magnitude of the Jahn-Teller
coupling.  

In comparison with the manual TB fits presented in
Ref.~\onlinecite{2007_ederer}, the construction of MLWFs is less
biased and more universally applicable. It allows to \emph{calculate}
parameters of the model instead of \emph{fitting} them to either
experimental or computational data. In particular, it is possible to
obtain accurate TB representations even for rather complex band
structures. However, care has to be applied when parameters
corresponding to a more complex parameterization are used for simpler
models. For example, using the MLWF nearest neighbor hopping
amplitudes within a simple model that neglects all further neighbor
hoppings, can lead to a significant underestimation of the total
bandwidth, so that in certain cases a parameterization with
renormalized nearest neighbor hoppings, leading to a more accurate
total bandwidth, might be more desirable. The analysis presented in
this work demonstrates that, depending on the specific application at
hand, MLWFs can in principle be used to construct more and more
refined TB parameterizations which lead to realistic,
materials-specific band structures with very high accuracy.

\begin{acknowledgments}
  This work was supported by Science Foundation Ireland under
  Ref.~SFI-07/YI2/I1051 and made use of computational facilities
  provided by the Trinity Center for High Performance Computing.
\end{acknowledgments}

\bibliography{references}

\begin{thebibliography}{26}
\bibitem[{foo({\natexlab{a}})}]{footnote:k-points}
\bibinfo{note}{Here and in the following we use the notation of
  Ref.~\onlinecite{Bradley/Cracknell:Book} to denote special k-points.}
\bibitem[{foo({\natexlab{b}})}]{footnote:eg-t2g}
\bibinfo{note}{Note that we are generally referring to
  \mbox{$|3z^3-r^2\rangle$} and \mbox{$|x^2-y^2\rangle$} as ``$e_g$-orbitals''
  and to all other $d$ orbitals as ``$t_{2g}$-orbitals'', even in cases where,
  strictly speaking, these are not the correct symmetry labels.}
\end{thebibliography}

\end{document}